\documentclass[twocolumn]{aastex63}

\usepackage{bm}
\usepackage{soul}
\usepackage{amsmath}
\usepackage{enumerate}
\usepackage[caption=false]{subfig}

\newcommand{\Msun}{\mathrm{M}_{\odot}}
\definecolor{darkgreen}{rgb}{0,0.8,0}

\submitjournal{ApJ}

\shorttitle{Alignment of an IMBH with a Stellar Disk}

\shortauthors{Sz\"olgy\'en et al.}

\begin{document}

\title{Resonant Dynamical Friction in Nuclear Star Clusters: \\ 
Rapid Alignment of an Intermediate-mass Black Hole with a Stellar Disk}

\correspondingauthor{\'Akos Sz\"olgy\'en}
\email{akos.szolgyen@ttk.elte.hu}

\author[0000-0001-6062-2694]{\'Akos Sz\"olgy\'en}
\affil{Institute of Physics, E\"otv\"os University,
P\'azm\'any P. s. 1/A, Budapest, 1117, Hungary}

\author[0000-0001-7982-6064]{Gergely M\'ath\'e}
\affil{Institute of Physics, E\"otv\"os University,
P\'azm\'any P. s. 1/A, Budapest, 1117, Hungary}

\author[0000-0002-4865-7517]{Bence Kocsis}
\affil{Rudolf Peierls Center for Theoretical Physics, University of Oxford,
Parks Road, Oxford, OX1 3PU, United Kingdom}

\begin{abstract}
We investigate the dynamical evolution of an intermediate-mass black hole (IMBH) in a nuclear star cluster hosting a supermassive black hole (SMBH) and both a spherical and a flattened disk-like distribution of stellar-mass objects. We use a direct N-body ($\varphi$\textsc{GPU}) and an orbit-averaged (\textsc{N-ring}) numerical integrator to simulate the orbital evolution of stars and the IMBH. We find that the IMBH's orbit gradually aligns with the stellar disk if their mutual initial inclination is less than 90$^\circ$. If it is larger than 90$^\circ$, i.e. counterrotating, the IMBH does not align. Initially, the rate of orbital reorientation increases linearly with the ratio of the mass of the IMBH over the SMBH mass and it is orders of magnitude faster than ordinary (i.e.~Chandrasekhar) dynamical friction, particularly for high SMBH masses. The semimajor axes of the IMBH and the stars are approximately conserved. This suggests that the alignment is predominantly driven by orbit-averaged gravitational torques of the stars, a process which may be called \textit{resonant dynamical friction}. The stellar disk is warped by the IMBH, and ultimately increases its thickness. This process may offer a test for the viability of IMBH candidates in the Galactic Center. Resonant dynamical friction is not limited to IMBHs; any object much more massive than disk particles may ultimately align with the disk. This may have implications for the formation and evolution of black hole disks in dense stellar systems and gravitational wave source populations for LIGO, VIRGO, KAGRA, and LISA. 
\end{abstract}


\section{Introduction} 
The stellar distribution in the Milky Way's nuclear star cluster shows intriguing dynamical behavior around the central SMBH, Sgr A$^\star$. A young coeval population of massive  Wolf\---Rayet stars, luminous blue variables and O-type stars are observed in one or two warped disks in the $0.03-0.5$ pc vicinity of Sgr A$^\star$ \citep{Schodel2005,Bartko2009,Bartko+2010,Lu2009,Yelda2014,Schodel+2018,Gallego-Cano+2018,Ali2020}; some of these massive stars form a tightly bound group (e.g.~IRS 13E complex) \citep{Schodel2005}, while old low-mass stars are observed to be spherically distributed \citep{Feldmeier2014,Schodel+2018,Gallego-Cano+2018}. Two main formation channels have been proposed to explain the observed distribution \citep{Tremaine1975,Milosavljevic2001,Hansen_2003,Kim_2004,Levin_Wu_Thommes2005,Antonini2012,Antonini2013,Alig2013,Mapelli+2013,Gnedin2014,Antonini2014,ArcaSedda2014,Antonini2015,ArcaSedda2015,ArcaSedda2017,Tsatsi2017,ArcaSedda2018,Trani2018,ArcaSedda2019,Mastrobuono2019,Schodel2020,ArcaSedda+2020,Do+2020}: the observed distribution is either (i) a remnant of a previous star-forming gas disk; or (ii) the stars have been delivered via massive compact star clusters from the surrounding regions of the Galaxy. 

Both scenarios may lead to the formation or delivery of IMBHs in the galactic nucleus \citep{Goodman_Tan2004,PortegiesZwart+2006,McKernan+2012,McKernan+2014,Fragione2018_b,Arca-Sedda_Gualandris2018,Askar+2021, arca_sedda_2019, 10.1093/mnras/sty3096,FragioneSilk2020} where they can play an important role in the local dynamics \citep{Gualandris_2009} triggering star formation \citep{Thompson+2005}, accelerating two-body relaxation \citep{Mastrobuono-Battisti+2014}, disrupting binaries \citep{Deme+2020}, leading to tidal disruption events \citep{Chen+2009,Chen_Lu2013,Fragione2018_a} and extreme mass ratio inspirals \citep{Bode2014}. The detection of IMBHs in the Galactic Center may become possible using precision astrometry \citep{Gualandris+2010,Girma2019,Naoz+2020} or pulsar timing \citep{Kocsis_Ray_PortegiesZwart2012}. Recently, promising IMBH candidates have been identified in the centers of compact gas clumps near the Galactic Center, supporting formation scenario (i): CO–0.40–0.22 \citep{Oka2017,Ballone2018}, HCN–0.009–0.044 \citep{Takekawa2019}, CO–0.31+0.11 \citep{Takekawa2019b} and HCN-0.085-0.094 \citep{Takekawa2020}. However, IRS 13E complex can be an example of scenario (ii) for the delivery of an IMBH to the Galactic Center (\citealt{Schodel2005,Tsuboi2017,Zhu2020,Greene2020}, but see \citealt{Petts_Gualandris2017}). The observed metal-poor rotating stellar subpopulation implies that at least $7\%$ of the nuclear star cluster may have been delivered by scenario (ii) \citep{ArcaSedda+2020,Do+2020}.

The IMBHs are expected to sink deep into the Galactic Center due to dynamical friction \citep{Levin_Wu_Thommes2005,Mastrobuono-Battisti+2014}. In the vicinity of the central SMBH, the IMBHs settle on short-period ($1-10^4$ yr) approximately Keplerian orbits. Then the eccentricity and the orbital inclination change predominantly due to the orbit-averaged gravitational interactions of the stars, a process known as resonant relaxation \citep{Rauch1996}. Since the mean spherical potential of the nuclear star cluster also drives rapid apsidal in-plane precession with $10^4$--$10^5$ yr period, this limits the resonant accumulation of torques that drive the eccentricity evolution. However, one component of the resonant torques that drive the reorientation of the orbital planes is not limited by apsidal precision; this process is known as vector resonant relaxation (VRR). 

The timescale of VRR has been determined for single-mass and two-component clusters of stellar objects to be between $10^5-10^7$ yr \citep{Hopman2006,Eilon2009,Kocsis2011,Kocsis2015,Giral2020}. This timescale is sufficiently short that the orbital inclinations may be expected to reach a quasi-stationary equilibrium distribution while the eccentricities and semimajor axes change much more slowly \citep{Roupas2017,Takacs2018,Bar-Or_Fouvry2018,Fouvry2019,Roupas2020}. However, entropy is expected to be maximized when the inclination distribution exhibits mass segregation, meaning that higher-mass objects are ultimately confined to smaller root-mean-square inclinations compared to lower-mass objects \citep{Rauch1996,Roupas2017,Szolgyen2018,Gruzinov+2020}. The eccentricity of higher-mass objects may also become systematically different \citep{Bar-Or_Fouvry2018,Gruzinov+2020}. The systematic change of the eccentricity of massive objects has been confirmed by numerical simulations \citep{Levin_Wu_Thommes2005,Alexander+2007,Lockmann+2009,Iwasawa2011,Sesana2011,Madigan_Levin2012,Foote+2020,Bonetti2020}. Mass segregation in the vertical direction has also been identified in numerical simulations of initially strongly non-axisymmetric nuclear stellar disks that lack a spherical cusp of stars \citep{Foote+2020}.

In multi-mass gravitating systems, dynamical friction drives the relaxation toward a mass-dependent statistical equilibrium. In such a system, a massive object may ``polarize'' the medium and the perturbations get amplified by collective gravitational effects, which backreact and lead to a rapid relaxation toward a statistical equilibrium \citep{Sellwood2013,Fouvry+2015,Fouvry_Pichon_Magorrian2017,Sridhar_Touma2017,Lau_Binney2019,Hamilton+2018,Hamilton2020,Hamilton2021,Fouvry+2021}. This may lead to a resonantly enhanced rate of dynamical friction if the mean-field potential admits action-angle variables \citep{Lynden-Bell_Kalnajs1972,Tremaine_Weinberg1984,Weinberg1989,Nelson_Tremaine1999,Chavanis2012,Heyvaerts+2017,Fouvry_Bar-Or2018,Bortolas+2020,Bortolas+2021,Banik_vandenBosch2021} which may result in orders-of-magnitude faster relaxation than predicted by Chandrasekhar's estimate \citep[however, see][]{Inoue2011,Petts+2016}.

In this paper, we examine the dynamical mechanism which leads to the rapid reorientation of the orbital plane of a massive object (e.g.~an IMBH) in response to a population of lower-mass stars orbiting a SMBH in a disk configuration, a process that may be called \textit{resonant dynamical friction} (RDF) \citep{Rauch1996}. In particular, we investigate how an IMBH settles into the midplane of a stellar disk, such as the clockwise disk around Sgr A$^\star$ in the Galactic Center  \citep{Bartko2009,Lu2009,Yelda2014,Gillessen2017}, and study the response of the disk using numerical simulations. 

We use two different numerical methods: (i) $\varphi$\textsc{GPU} \--- a direct N-body simulator accelerated by graphical processing units \citep{Berczik+2011,Berczik+2013}, and (ii) \textsc{N-ring} \--- a secular N-body simulator that uses the orbital- and precession-period averaged interactions \citep{Kocsis2015}.Both $\varphi$\textsc{GPU} and \textsc{N-ring} account for the superposition of the interactions between all pairs of particles, but while $\varphi$\textsc{GPU} simulates the instantaneous interaction between the point masses \textsc{N-ring} calculates the interaction between annuli covered during the orbital and precession period. The semimajor axes and eccentricities of different stars are free to change in $\varphi$\textsc{GPU} while they are fixed by construction in \textsc{N-ring}. The difference between the two methods allows us to identify the main dynamical mechanism driving the evolution and in particular to explore the contribution of Chandrasekhar's (ordinary nonresonant) dynamical friction (CDF) and resonant dynamical friction (RDF) to the process of inclination relaxation. We find that the \textsc{N-ring} simulations, which account for RDF but not CDF by construction, match the results of the $\varphi$\textsc{GPU} simulations that include both CDF and RDF. The alignment process is much more rapid in the simulations than expected by a simple analytic estimate of CDF. We examine the process using numerical simulations with different initial IMBH orbital inclinations, IMBH masses, number of stars, and radial surface density profiles for the stellar disk. We measure the warp and the thickness of the disk, as well as the evolution of the IMBH's semimajor axis, eccentricity, and inclination. We construct a simple empirical analytic model for the alignment via RDF. 

\section{Nonresonant dynamical friction} \label{sec:nonresonant}
We start with a simple analytic estimate of the alignment time of the IMBH due to Chandrasekhar's dynamical friction (CDF) with respect to the disk stars during disk crossings. The mean deceleration of the IMBH is given by Chandrasekhar's formula for a homogeneous medium as \citep[Eq. 8.1a in][]{Binney2008}:

\begin{align}\label{eq:dynamicalfriction}
\frac{\mathrm{d}\mathbf{v}_\mathrm{IMBH}}{\mathrm{d}t} = 4 \pi G^2 & m_\mathrm{IMBH} \rho \ln(\Lambda)
    \int{\mathrm{d}^3 \mathbf{v} f(\mathbf{v}) \frac{\mathbf{v}-\mathbf{v}_\mathrm{IMBH}}{| \mathbf{v} - \mathbf{v}_\mathrm{IMBH}|^3}  }
\end{align}

where $\rho$ is the density of stars and $\ln(\Lambda)$ is the Coulomb logarithm with $\Lambda = h_\mathrm{d}/b_{90}$. $b_{90} = G (m_\mathrm{IMBH} + m)/\langle v_\mathrm{rel}^2 \rangle \ll  h_\mathrm{d}$, where $v_\mathrm{rel}$ is the relative velocity of stars and the IMBH, $h_\mathrm{d}$ is the thickness of the disk, and $f(\mathbf{v})$ is the probability density function of disk star velocities in a small box around the crossing point of the IMBH.\footnote{To calculate $b_{90}$, note that we assume that $h_\mathrm{d}$ is smaller than the Hill radius $r_\mathrm{H}=(m_\mathrm{IMBH}/m_\mathrm{SMBH})^{1/3} r = 0.1\,r$, which holds for $h_\mathrm{d}=0.01\,$pc  and $r>0.1\,$pc in our fiducial model.} The alignment time due to dynamical friction accounting for two disk crossings of duration $t_{\rm cross}=h_{\rm d}/v_{\mathrm{IMBH},z}$ per orbit of period $t_{\rm orb}$ may be estimated as
\begin{equation}
t_\mathrm{align,CDF}=\left(\dfrac{\rm{d}\iota}{\rm{d}t}\right)^{-1}=\frac{t_{\rm orb}}{2 t_\mathrm{cross}} \frac{v_{\mathrm{IMBH},z}}{a_{\mathrm{CDF},z}} 
\end{equation}
where $v_{\mathrm{IMBH},z}$ is the IMBH's velocity in the $z$-direction, $a_{\mathrm{CDF},z} = \mathrm{d}v_{\mathrm{IMBH},z} / \mathrm{d}t$ and $\iota$ is the inclination angle of the IMBH with respect to the stellar disk.

In the limit of small IMBH orbital eccentricity and a cold thin disk of total mass $m_{\rm d}$ and maximum radius of $r_{\rm d}$ with density 
\begin{equation}\label{eq:rho}
 \rho(r,z) = \frac{2-\gamma}{2\pi r_{\rm d}^2 h_{\rm d}} \left(\frac{r}{r_d}\right)^{-\gamma}m_{\rm d}   ~~\mathrm{if}~ |z|<\frac{h_{\rm d}}{2}
\end{equation}
and $\rho(r,z)=0$ otherwise. We define Cartesian coordinates $(x,y,z)$ where the disk plane is spanned by $\{x,y\}$ and the IMBH crosses the $z=0$ midplane at $x=0$,
substitute $\mathbf{v}_{\rm IMBH}= (v_{\rm K}\cos \iota_0,0,v_{\rm K}\sin\iota_0)$ and assume that all stars have $\mathbf{v}=(v_{\rm K},0,0)$ in the vicinity of the crossing point of the IMBH in the disk in Eq.~\eqref{eq:dynamicalfriction} where $v_{\rm K}=(Gm_{\rm SMBH}/r)^{1/2}$ is the Keplerian velocity. Then the alignment time simplifies to 
\begin{align}\label{eq:talign}
 t_{\rm align,CDF} &= \frac{2\sin\iota \sin^3(\iota/2)}{(2-\gamma)\ln\Lambda} 
  \left(\frac{r}{r_{\rm d}}\right)^{\gamma-2} 
 \frac{m_{\rm SMBH}^2}{m_{\rm d}m_{\rm IMBH}}
 t_{\rm orb}\nonumber
\\ 
 &=
 \frac{2\sin\iota\sin^3(\iota/2)}{\ln\Lambda} 
 \frac{m_{\rm SMBH}^2}{m_{\rm d,loc}m_{\rm IMBH}}
 t_{\rm orb}\,,
\end{align}
where we have introduced the ``local disk mass at $r$'' as
\begin{equation}\label{eq:mloc}
    m_{\rm d,loc} = \frac{\rm{d} m}{\rm{d} \ln r} = \int  2\pi r^2 \rho\, \rm{d} z =   (2-\gamma) \left(\frac{r}{r_d}\right)^{2-\gamma}m_{\rm d}\,.
\end{equation}

The CDF alignment time is $43$ Myr for $\iota=45^{\circ}$ at $r=0.15\,$pc for our fiducial model with $m_{\rm SMBH}=10^6\Msun$, $m_{\rm IMBH}=10^3\Msun$, $m_{\rm d}=8,191\Msun$, $m_{\rm d,loc}=2,273\Msun$, $\gamma=0.75$, and $r_{\rm d}=0.5\,$pc. For our alternative model with $\gamma=1.75$ and all other parameters fixed, we have $m_{\rm d,loc}=1,516\Msun$, so $t_{\rm align,CDF} = 65\,\mathrm{Myr}$.
Note that the result depends on the radius and the stellar disk properties only through the orbital period and the local disk mass. It is otherwise independent of $h_{\rm d}$ and the mass of individual disk stars, but it shows a strong inclination dependence as shown in Figure \ref{fig:empirical-fit} below. 

We also find a similar result by measuring the CDF alignment time in the numerical simulations by substituting the actual velocity distribution of stars (Figure \ref{fig:velocity-distribution}, see Appendix~\ref{sec:appendix}) around the IMBH within a cylinder around the crossing point of the IMBH in the disk ($r=0.15\,$pc) of height $h_\mathrm{d}= 0.01 \mathrm{pc}$ (i.e. the local thickness of the stellar disk) and radius chosen arbitrarily to be $r_\mathrm{cyl} = 0.05\mathrm{pc}$ in Eq.~\eqref{eq:dynamicalfriction} and find a similar result. 

We note, however, that $t_\mathrm{align,CDF}$ is an upper limit for the true alignment time for, several reasons. First, dynamical friction is enhanced significantly in an inhomogeneous cuspy density profile around SMBH \citep{Arca-Sedda_Capuzzo-Dolcetta2014}. Further, dynamical friction may be accelerated by orbit-averaged torques, which drives resonant relaxation \citep{Madigan_Levin2012}. We confirm this estimate in detail below. For the system parameters listed above, the apsidal precession rate of the IMBH is $0.019$ times the orbital angular velocity around the SMBH according to Eq.~(A1) in \cite{Kocsis2015}. Thus, the apsidal precession period of the IMBH is $t_\mathrm{a,prec} = 0.28$ Myr, which is much larger than the orbital period, $t_\mathrm{orb} = 5.4$ kyr. However, it is much shorter than the initial nodal precession period of the IMBH and the stellar disk measured using \textsc{N-ring} to be $t_\mathrm{n,prec} =2.9$ Myr, which approximately characterizes the RDF timescale.\footnote{Note that, given the large initial inclination, the measured nodal precession timescale is consistent with the quadrupole approximation \citep{Nayakshin2005a,Kocsis2015} as $t_\mathrm{n,prec} \sim \frac{4}{3}(\cos\iota)^{-1} m_{\rm SMBH}(m_{\rm IMBH}+m_{\rm d,loc})^{-1}t_{\rm orb} = 3.1$ Myr} Eq.~\eqref{eq:talign} shows that $t_{\rm align,CDF}\propto \iota^4$ for small $\iota$, implying that CDF may play an important role in aligning the orbit during the later phase of the evolution when the inclination is small (see Section \ref{sec:analyticRDF} for further discussion).\footnote{Note that the $\iota^4$ scaling is valid only for small orbital inclinations but larger than the thickness of the disk.}

\section{Numerical Methods}
We run simulations using $\varphi$\textsc{GPU}  \citep{Berczik+2011,Berczik+2013,Li2012,Li2017}
a direct-summation $N$-body code that uses the Hermite integration scheme with individual block time steps to solve the instantaneous equations of motion of point particles with gravitational softening. We also run \textsc{N-ring} \citep{Kocsis2015} with the same initial conditions. In contrast with $\varphi$\textsc{GPU}, \textsc{N-ring} is a secular code that integrates the pairwise interactions of orbit-averaged stellar trajectories using a time-reversible symplectic parallel scheme using a multipole expansion. By construction, \textsc{N-ring} neglects variations in the semimajor axes and eccentricities, implying that it neglects two-body relaxation and scalar resonant relaxation, and thus simulates the effects of VRR only, while $\varphi$\textsc{GPU} does not make any such approximations. We direct the readers to the references for a detailed discussion of these numerical methods. \textsc{N-ring} neglects the contribution of multipoles with $\ell>50$ which amounts to softening with an angular separation of $1.5^{\circ}$ in angular momentum direction space, while $\varphi$\textsc{GPU} applies softening with $10^{-5}$ pc for stars and $10^{-3}$ pc for the SMBH and IMBH. We use both codes to understand the relative contribution of VRR and two-body relaxation to the inclination relaxation. 

First, we run 6 $\varphi$\textsc{GPU} and \textsc{N-ring} simulations with different initial inclinations for the IMBH's orbit as specified below. Then, we explore the evolution for different disk surface density profiles and IMBH masses using \textsc{N-ring} to determine the main properties of resonant dynamical friction and to derive an empirical analytic model.

\subsection{Initial conditions} \label{sec:initial}
Our models consist of a disk of $N = 8191$ equal-mass $1\,\Msun$ stars together with an $m_\mathrm{IMBH}=10^3\,\Msun$ IMBH and a $m_\mathrm{SMBH}=10^6\,\Msun$ SMBH fixed at the center. In addition, we include a static Plummer-potential $\Phi_\mathrm{P} (r) = - G m_\mathrm{P} / \sqrt{r^2 + r^2_\mathrm{P}}$ with $m_\mathrm{P}=10^5 \Msun$ and $r_\mathrm{P} = 0.2$ pc, in order to mimic the gravity of the spherical stellar component in the galactic nucleus, which drives apsidal precession and quenches the coherent torques of eccentric orbits.\footnote{Without the spherical component, the disk may preserve a strongly non-axisymmetric structure, if present initially, which may strongly influence the evolution (\citealt{Madigan+2009,Madigan_Levin2012,Madigan+2018,Foote+2020,Rodriguez+2021}, but see \citealt{Gualandris+2012}). A possibly important simplification for the spherical component of mass distribution in our simulations is the assumption of a smooth mean-field potential, which neglects $N^{1/2}$ fluctuations caused by the finite number of stars in the spherical cusp, which may excite long-wavelength warps in the disk \citep{Kocsis2011} and also may affect the evolution of the IMBH. We leave the study of the influence of the fluctuations of the spherical component to a future study.} While this model is obviously far from a realistic representation of the Galactic Center, which has a Bahcall-Wolf cusp of $10^7$ stars with a possible inner cavity, a population of S-stars, a warped and twisted stellar disk of $10^4\,\Msun$ stars, $10^4$ black holes, a massive molecular torus, etc. \citep{Genzel2010}, it serves as a simplified model to gain an understanding of the orbital alignment of a massive point mass in an anisotropic system. 

The stars are initially distributed in the simulation in an axisymmetric thin disk with height $h_{\rm d}=0.01$ pc, which is uniform in cylindrical coordinates along the azimuth angle and the $z$-axis. The surface density in our fiducial model is chosen according to Eq.~\eqref{eq:rho} with surface density $\propto r^{-0.75}$ between $0.03 - 0.5$ pc from the SMBH. This matches the expectation for the cluster infall scenario \citep{Berukoff2006}, but it is much shallower than the observed surface density profile of the clockwise disk in the Galactic Center, which is proportional to $r^{-2.1\pm0.4}$ \citep{Lu2009}, $r^{-1.95\pm0.25}$ \citep{Bartko2009}, or $r^{-1.4\pm0.2}$ \citep{Bartko+2010}. We also explore a steeper density profile of $r^{-1.75}$ separately below. The initial velocities of stars are perturbed with respect to the circular velocities\footnote{Note that $v_{c,i}$ is the Keplerian circular velocity for a point mass of $m_\mathrm{SMBH}+m_\mathrm{P}$, which is somewhat larger than the circular velocity of the actual SMBH+Plummer potential by a factor  $(m_\mathrm{SMBH}+m_\mathrm{P})/\{m_\mathrm{SMBH}+m_\mathrm{P}[1+(r_{\rm P}/r)^2]^{-3/2}\}\approx 1 + (m_\mathrm{P}/m_\mathrm{SMBH})\{1-[1+(r_{\rm P}/r)^2]^{-3/2}\}$, which is between 1 and 1.1.} corresponding to the SMBH and the Plummer-potential component: $ v_i = v_{c,i} \pm 0.05 f(v_{c,i}) $, where $v_{c,i}^2 = G (m_\mathrm{SMBH}+m_\mathrm{P})/\sqrt{x_i^2 + y_i^2}$ is the $i$th star's circular velocity, $f(v_{c,i})$ is a randomly sampled value from the PDF of a Maxwell-Boltzmann distribution that has the mean at $v_{c,i}$, and $x_i$, $y_i$ are the coordinates of the $i$th star. We choose the direction of $v_{c,i}$ such that $95\%$ and $5\%$ of stars are orbiting respectively in the same sense and in the opposite sense as the IMBH in the disk. In comparison, 19 out of 90 WR/O-stars in the Galactic Center are consistent with belonging to a retrograde disk \citep{Bartko2009} although they are also consistent with belonging to a spherical distribution \citep{Lu2009,Yelda2014}. This disk configuration is not in statistical equilibrium initially, thus we first let the system to evolve for a few million years until the disk reaches a mean-field equilibrium in which the inclination distributions of the inner, overlapping, and outer stars have mean and standard deviation $\langle \cos^2 \iota \rangle = 0.994,0.996,0.998$ and $\sigma(\cos^2 \iota ) = 0.022,0.036,0.012$, respectively; see Sec.~\ref{sec:thickness} below. We show the relaxed distribution of orbital parameters in the Appendix; see Figure \ref{fig:obital-elements}. 

We use this relaxed disk model together with an IMBH parameterized by $(m_\mathrm{IMBH},$ $a_\mathrm{IMBH},e_\mathrm{IMBH}) = (10^3\Msun,0.15\mathrm{pc},0.33)$ as the
initial condition in both the $\varphi$\textsc{GPU} and in the \textsc{N-ring} simulations. We run $6$ simulations with $\iota_0 =35^\circ,45^\circ,55^\circ,65^\circ,75^\circ,135^\circ$ different orbital inclinations for the IMBH. The initial orbital parameters and masses of stars, as well as the IMBH are identical in the $\varphi$\textsc{GPU} and the \textsc{N-ring} simulations. \textsc{N-ring} does not include the Plummer-potential component directly, but its influence is included in the assumption of rapid apsidal precession, as the interaction in \textsc{N-ring} is averaged over the orbital and the apsidal precession period.  

\subsection{Characterization of disk thickness and warp}\label{sec:thickness}
To characterize the state of the disk in the simulation, we introduce the quadrupole tensor of the stellar angular momenta: 
\begin{equation}\label{eq:Q}
Q_{\alpha\beta}=\dfrac{\sum_{i=1}^N L_{i\alpha}L_{i\beta}}{\sum_{i=1}^N{|\boldsymbol{L}_{i}|^2}},
\end{equation}
where $\boldsymbol{L}_{i}$ denotes the angular momentum vector of the $i^{\rm th}$ star, and $\alpha$ and $\beta$ respectively label the Cartesian vector components. The largest eigenvalue (denoted by $q$ hereafter) quantifies the flatness, as it is given by
\begin{equation}\label{eq:q}
    q = \langle \cos^2 \iota \rangle = \dfrac{\sum_{i=1}^N |\boldsymbol{L}_{i}|^2 \cos^2 \iota_i}{\sum_{i=1}^N |\boldsymbol{L}_{i}|^2}
\end{equation}
where $\iota_i$ is the orbital inclination of stars relative to the disk midplane, i.e., the angle subtended by $\boldsymbol{L}_{i}$ and the principal eigenvector of $Q_{\alpha\beta}$.\footnote{In particular, since $\mathrm{Tr\,}\boldsymbol{Q}=1$ and all eigenvalues are non-negative, the principal eigenvalue $q=1$ corresponds to the case where the other two eigenvalues are 0 and so all angular momentum vectors have $\iota=0^{\circ}$ or $180^{\circ}$ representing a razor-thin disk in physical space. Further, $q=1/3$ corresponds to the case where all eigenvalues are equal which represents an approximately isotropic spherical distribution of orbits both in angular momentum space and in physical space.}
\begin{figure*}
\centering 
\includegraphics[width=2.0\columnwidth]{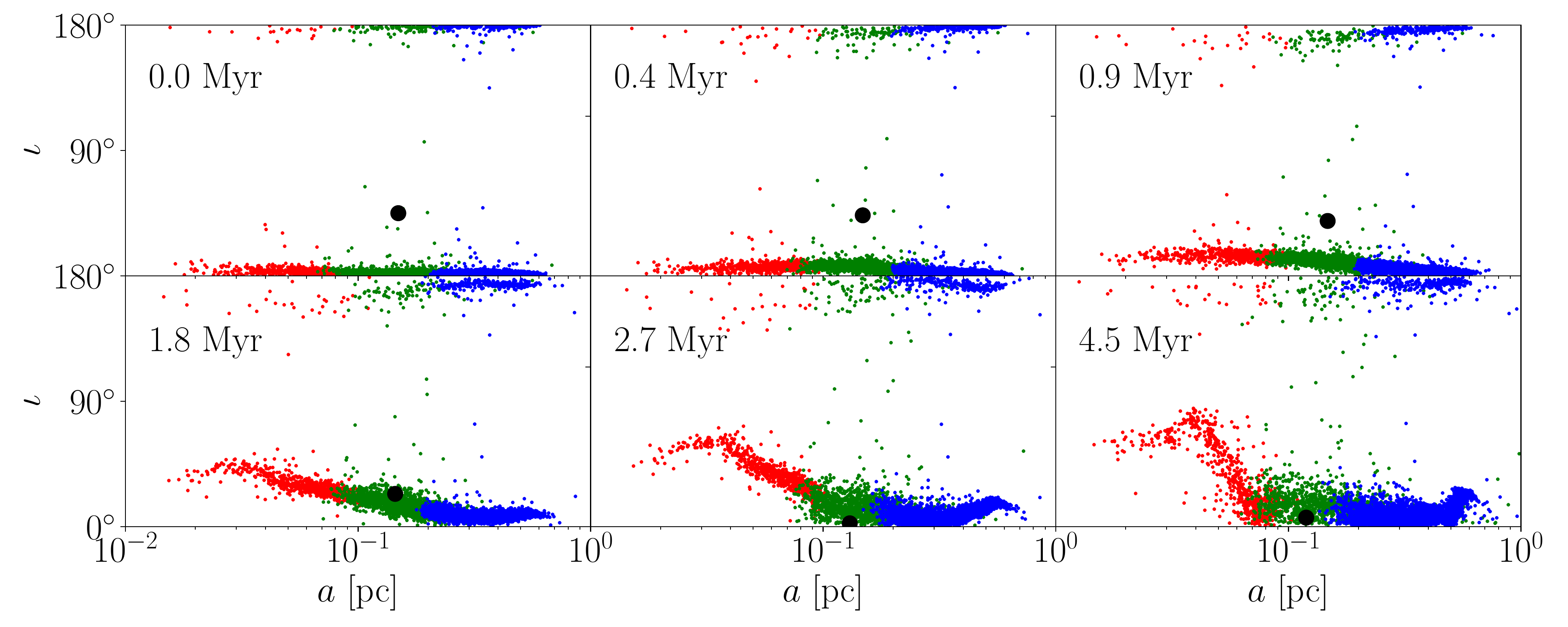}
\caption{\label{fig:snapshots} The inclinations and semimajor axes of stars (colored points) and the IMBH (black dot) in the $\varphi$\textsc{GPU} simulation shown as a scatter plot for 6 snapshots in the 6 panel at $0$, $0.4$, $0.9$, $1.8$, $2.7$, and $4.5$ Myr as labeled. Red, green, and blue points show the inner, overlapping, and outer region of the disk with respect to the IMBH orbit, respectively.  \href{http://galnuc.elte.hu/szolgyen_et_al_2021_Fig1.mp4}{See also the animated time evolution.}}
\end{figure*}

We characterize the root-mean-square warp angle or thickness of the disk by the quantity
\begin{equation}\label{eq:Di}
    \Delta \iota \equiv \cos^{-1} \sqrt{q}\,.
\end{equation}

Note that $\Delta \iota=0$ for a razor-thin flat disk, and $\Delta \iota$ is independent of an overall tilt of the disk but it increases with the warp of a thin disk and with the thickness. In the limit of an isotropic distribution, or a razor thin disk warped by $180^\circ$, $\Delta \iota = \cos^{-1} (3^{-1/2}) = 54.7^\circ$.

We measure $\Delta \iota$ for the inner, overlapping, and outer regions of the disk with respect to the IMBH, separately, by evaluating the sums over only the corresponding stars in Eq.~\eqref{eq:Q}. We define these regions depending on the peri- and apocenter $(r_p,r_a)$ of the stars with respect to the IMBH $(r_{p,\mathrm{IMBH}},r_{a,\mathrm{IMBH}})$ as
\begin{itemize}
    \item inner region: $r_a<r_{p,\mathrm{IMBH}}$,
    \item overlapping region: $r_p \leqslant r_{a,\mathrm{IMBH}}$ or $r_{p,\mathrm{IMBH}} \leqslant r_a$, 
    \item outer region: $r_{a,\mathrm{IMBH}}<r_p$.
\end{itemize}
The number of stars in the three regions are $\langle N_\mathrm{in} \rangle = 9.4\% \pm 0.64 \%$, $\langle N_\mathrm{overl} \rangle = 18.1\% \pm 4.3 \%$, and $\langle N_\mathrm{out} \rangle = 72.5\% \pm 4.0\%$, respectively, for our standard model. Here, the fluctuations are mostly due to variations in the IMBH eccentricity with time.

\section{Results}
We first present the results of $\varphi$\textsc{GPU} simulations for initial IMBH inclination $\iota_0 = 45^\circ$. Figure \ref{fig:snapshots} shows the scatter plot of orbital inclinations and semimajor axes of disk stars and the IMBH, $\cos \iota_i = \bm{L}_i\cdot \bm{L}_{\rm tot}/(|\bm{L}_i| |\bm{L}_{\rm tot}|)$, for 6 representative snapshots as indicated in the panels. Here $\bm{L}_{\rm tot}$ is the total angular momentum of the system. We also provide an animated video showing the complete structural evolution of the disk as the IMBH sinks down and aligns with the disk. The stars are depicted by red, blue, or green points depending on their peri- and apocenter $(r_p,r_a)$ with respect to the IMBH  $(r_{p,\mathrm{IMBH}},r_{a,\mathrm{IMBH}})$ (see definition below Eq.~\ref{eq:Di}): inner orbits in red; outer orbits in blue; and overlapping orbits in green. The black filled circle indicates the IMBH. The figure shows that initially the disk is flat and thin and the IMBH is at $\iota_0=45^{\circ}$. The inclination angles start to change visibly at $0.4$ Myr. At $0.9$ Myr, the inner regions develop a larger warp. At $1.8$ Myr, the IMBH aligns the disk first and the disk is significantly warped in the inner and overlapping region. At $2.7$ and $4.5$ Myr, the IMBH settles into the midplane, and the disk warp is limited mostly to the inner region to within $r<0.08\,\rm pc$, where the local disk mass Eq.~\eqref{eq:mloc} is smaller than the IMBH mass, but the disk thickness in the overlapping and the close by parts of the outer disk is much larger than initially. We examine the evolution of the orbital elements of the IMBH and the response of the disk in detail next.

\subsection{Semimajor-axis and eccentricity evolution}
\begin{figure}
\includegraphics[width=1\columnwidth]{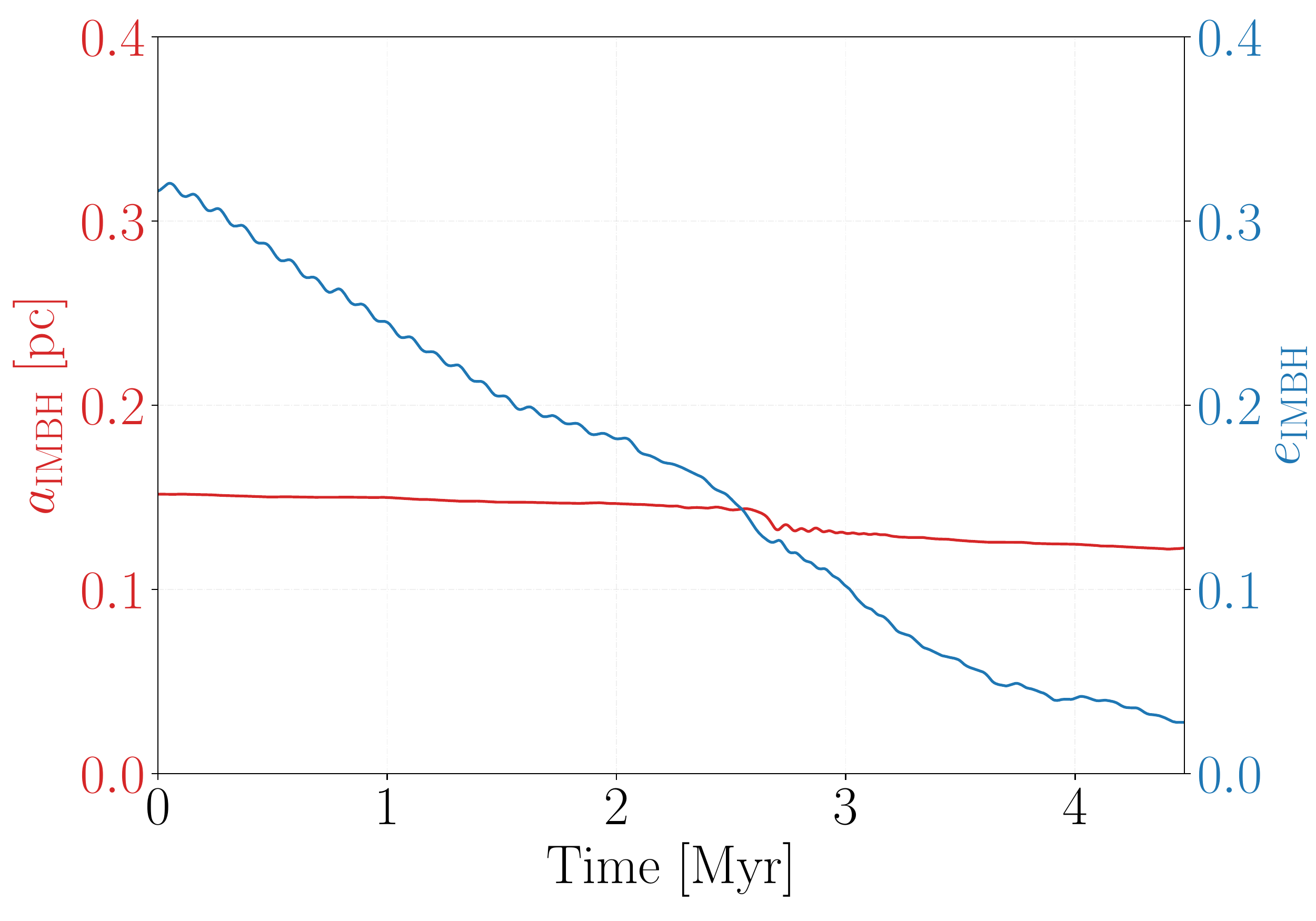}
\caption{\label{fig:a-e-t} The semimajor axis (red, left axis) and the eccentricity (blue, right axis) of the IMBH as a function of time. For the evolution of the distribution of semimajor axis and eccentricity of the stellar disk, see Figure~\ref{fig:obital-elements} in the Appendix.}
\end{figure}

Figure \ref{fig:a-e-t} shows the semimajor-axis and eccentricity evolution of the IMBH. The IMBH orbit circularizes during the relaxation process as also found by \citet{Madigan_Levin2012} for a corotating disk and by \citet{Bonetti2020}. The mean eccentricity decays from $0.33$ to $0.02$ during $4.5$ Myr. The semimajor axis is approximately constant during the evolution to within $6\%$ until the IMBH aligns with the disk at $2.6$ Myr, when the semimajor axis decreases suddenly by another $6\%$ within $0.1$ Myr then decreases gradually by $9\%$ until $4.5$ Myr.

The semimajor-axis distribution of stars shows a depletion around the IMBH's semimajor axis in the bottom left panel of Figure \ref{fig:obital-elements} in the Appendix. The scalar angular momentum of the IMBH varies by less than $6\%$.

\subsection{Orbital alignment and warps}
\begin{figure}
\centering 
\includegraphics[width=1.0\columnwidth]{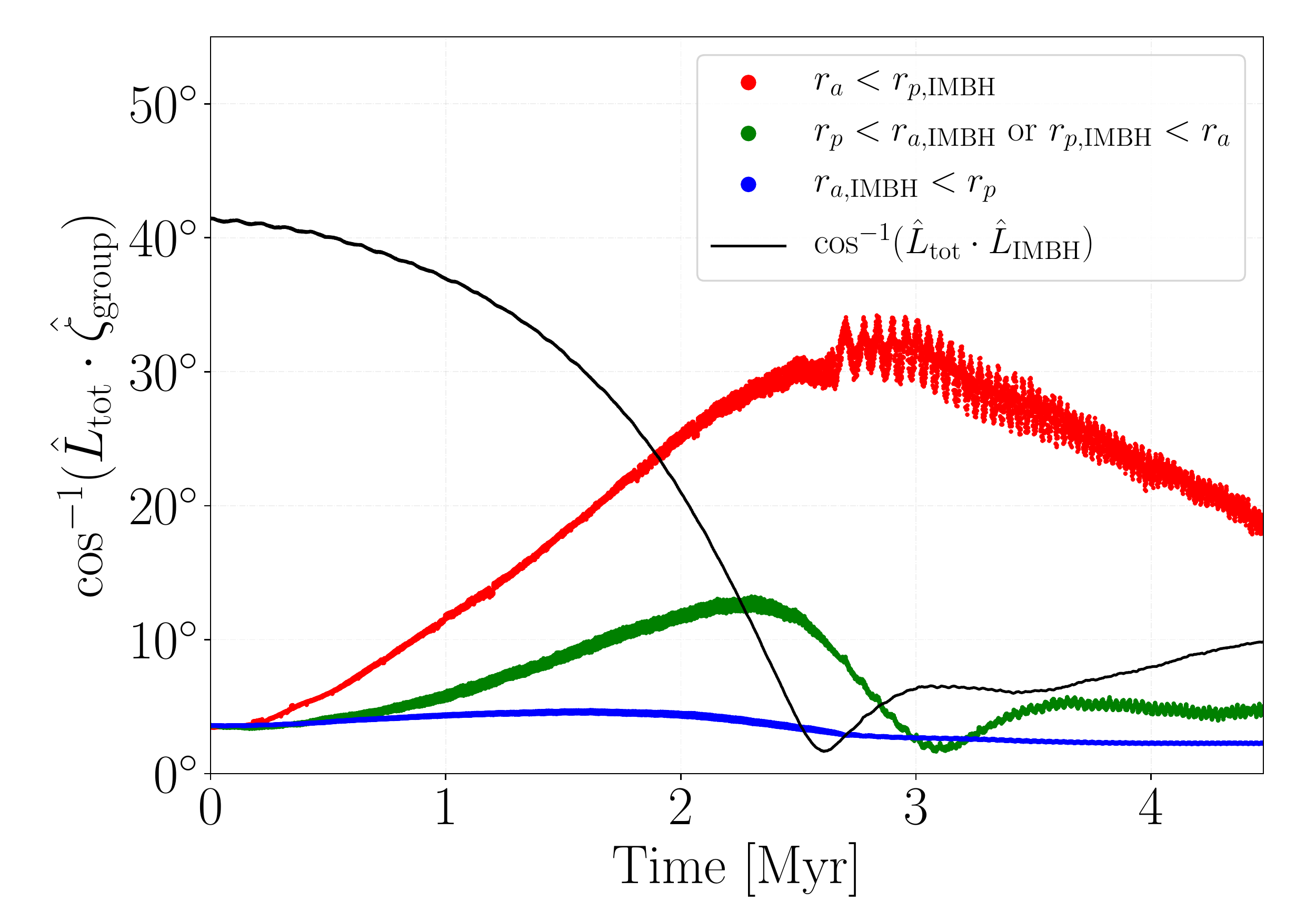}\\
\includegraphics[width=1.0\columnwidth]{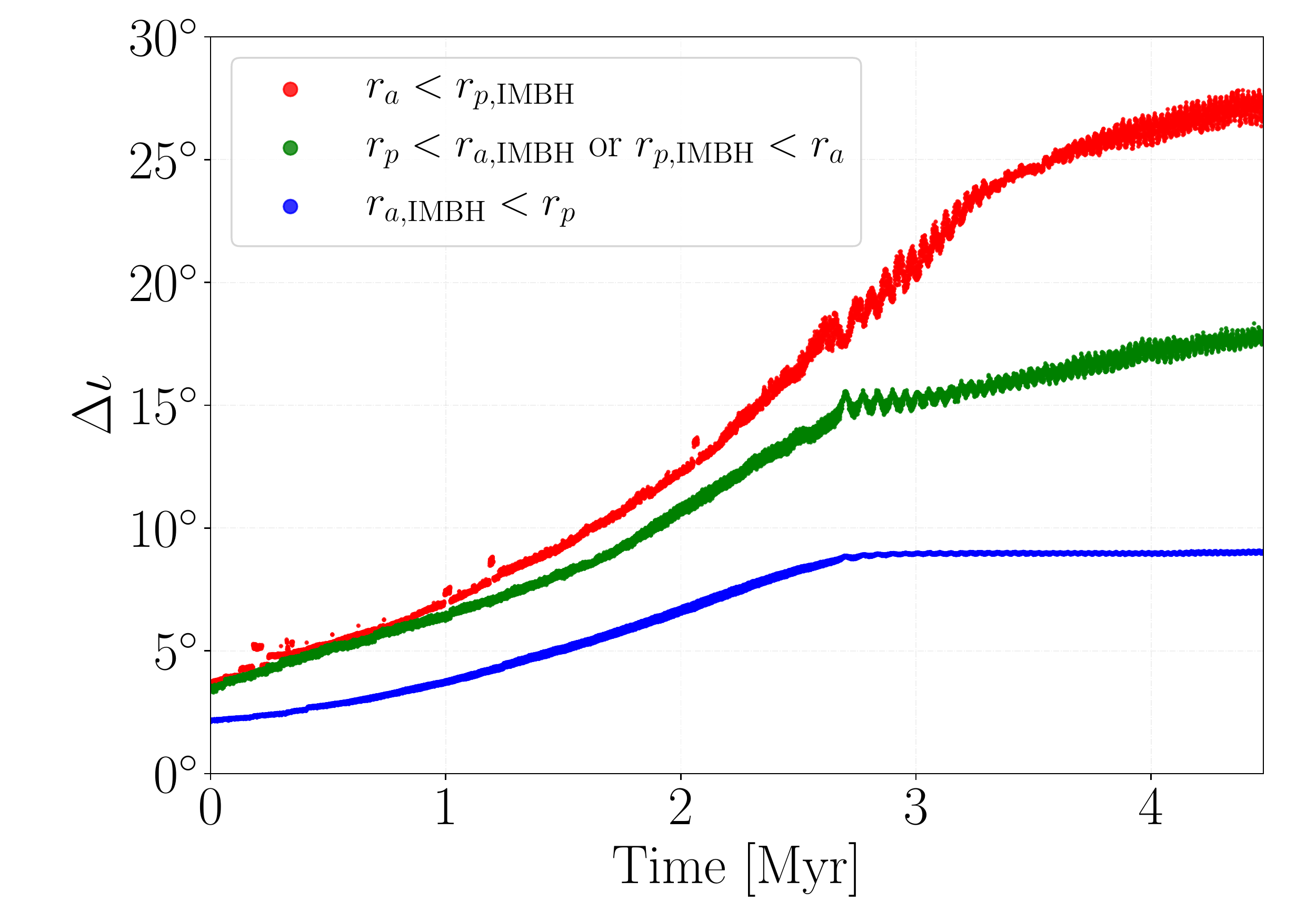}%
\caption{\label{fig:inclination-t} 
\textit{Top}:
The time evolution of the orbital inclination of the IMBH and stars for the inner, overlapping, and outer region with respect to the total angular momentum, $\hat{L}_\mathrm{tot}$. Black curve shows the alignment angle of the IMBH, defined as the angle between the angular momentum $\hat{L}_\mathrm{IMBH}$ and $\hat{L}_\mathrm{tot}$. We  characterize the orientation of the stellar disk with the respective principal angular momentum eigenvectors (Eq.~\ref{eq:Q}) of the stars in the given radial group, $\zeta_{\rm group}$. Red, green, and blue curves show the angle between $\hat{\zeta}_{\rm group}$ and $\hat{L}_\mathrm{tot}$ for the stars in the inner, radially overlapping, and outer parts of the disk with respect to the IMBH. \textit{Bottom:} The time evolution of the disk warp/thickness in the inner (red), overlapping (green), and outer region (blue) defined by Eq.~\eqref{eq:Di}.}
\end{figure}
The top panel in Figure~\ref{fig:inclination-t} shows the time evolution of the angle between $\bm{L}_{\rm IMBH}$ and $\bm{L}_{\rm tot}$ (black line), where $\bm{L}_{\rm IMBH}$ is the angular momentum vector of the IMBH. Colored curves show the angle between the respective principal angular momentum eigenvector\footnote{Note that the principal angular momentum eigenvector characterizes the instantaneous mean orientation/tilt of the disk; it is approximately parallel with the mean angular momentum of the given region to less than $5.2^\circ$ in our reference simulation.} of the disk stars (i.e. that of  $Q_{\alpha\beta}$, see Eq.~\ref{eq:Q}) and $\bm{L}_{\rm tot}$ separately for the inner, overlapping, and outer stars as defined below Eq.~\eqref{eq:Di}. The bottom panel shows $\Delta \iota$ (Eq.~\ref{eq:Di}). In practice, the top panel characterizes the evolution of an overall tilt of the three parts of the disk, and the bottom panel characterizes the evolution of the warp and/or the disk thickness within the three regions. The top panel shows that the IMBH sinks into the plane of the disk in an accelerated way, and settles after around $2.6$ Myr. The disk is initially flat, but the inner disk quickly becomes tilted as the orientation of the innermost part of the disk separates from that of the overlapping and the outer disk in the top panel. At around $>0.6$ Myr, the overlapping part of the disk also decouples from the outer disk. The inner disk tilt peaks at $\sim 30^\circ$ at $\sim 2.5$ Myr, while the overlapping stars' inclination peaks around $\sim 12^\circ$ at almost the same time. \footnote{The strong variation in the inner disk after $2.6$ Myr is a numerical artifact, as the particle representing the spherical stellar background distribution (i.e the Plummer potential) in the simulation gets displaced from the center by $0.005$ pc.}

The bottom panel of Figure \ref{fig:inclination-t} shows that the curvature or thickness of the disk increase in all three regions until the IMBH settles into the disk at $2.6$ Myr, and this is more prominent in the inner and overlapping regions. The thickness/curvature continues to increase only in the inner and overlapping regions after the IMBH aligns with the disk. By $4.5$ Myr, $\Delta \iota$ is increased by respective factors of $7.4$, $5.1$, and $4.1$ in the inner, overlapping, and outer regions. 

\subsection{Dependence on the initial IMBH inclination}
\begin{figure}
\centering 
  \includegraphics[width=1.0\columnwidth]{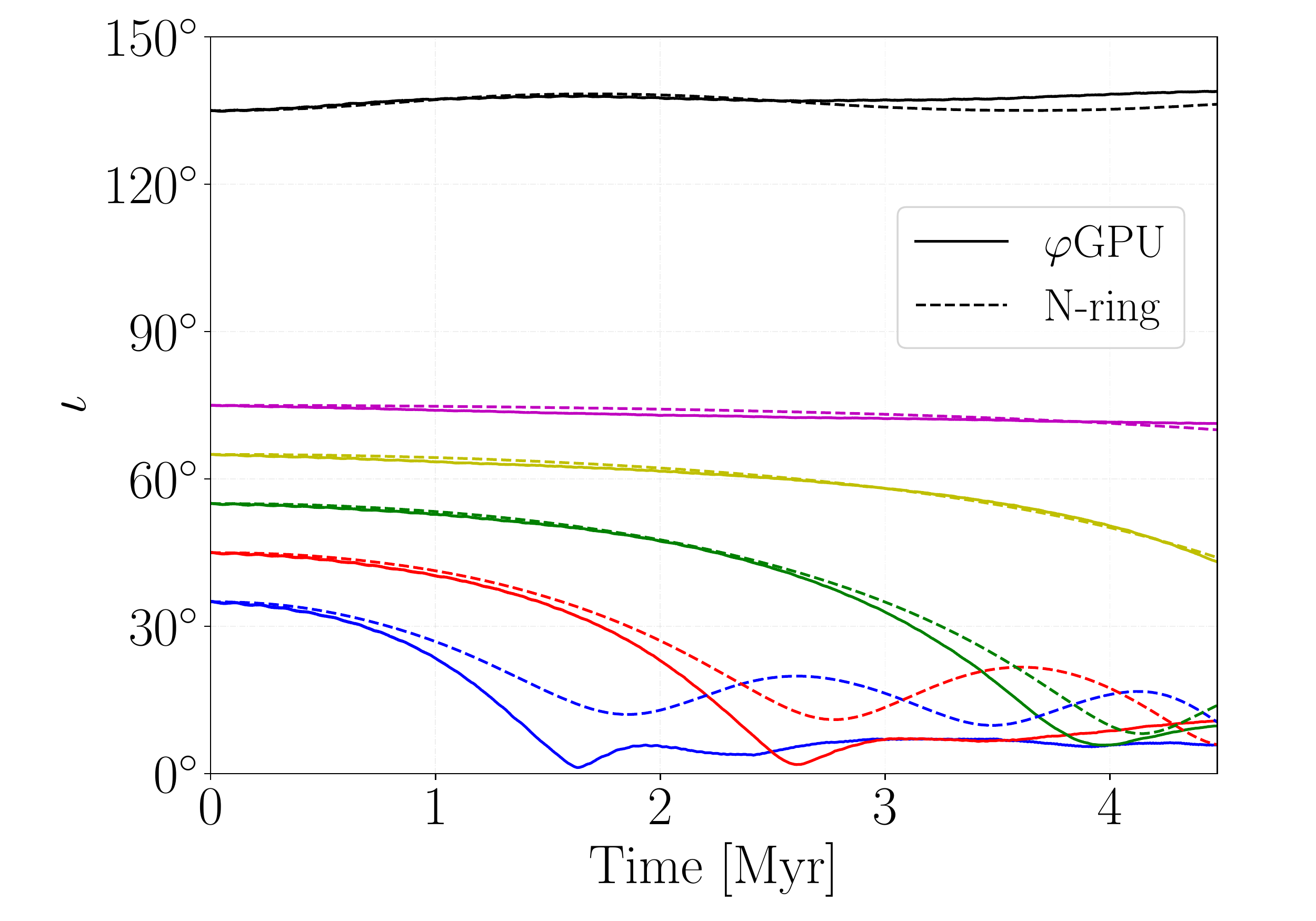}\\
\includegraphics[width=1.0\columnwidth]{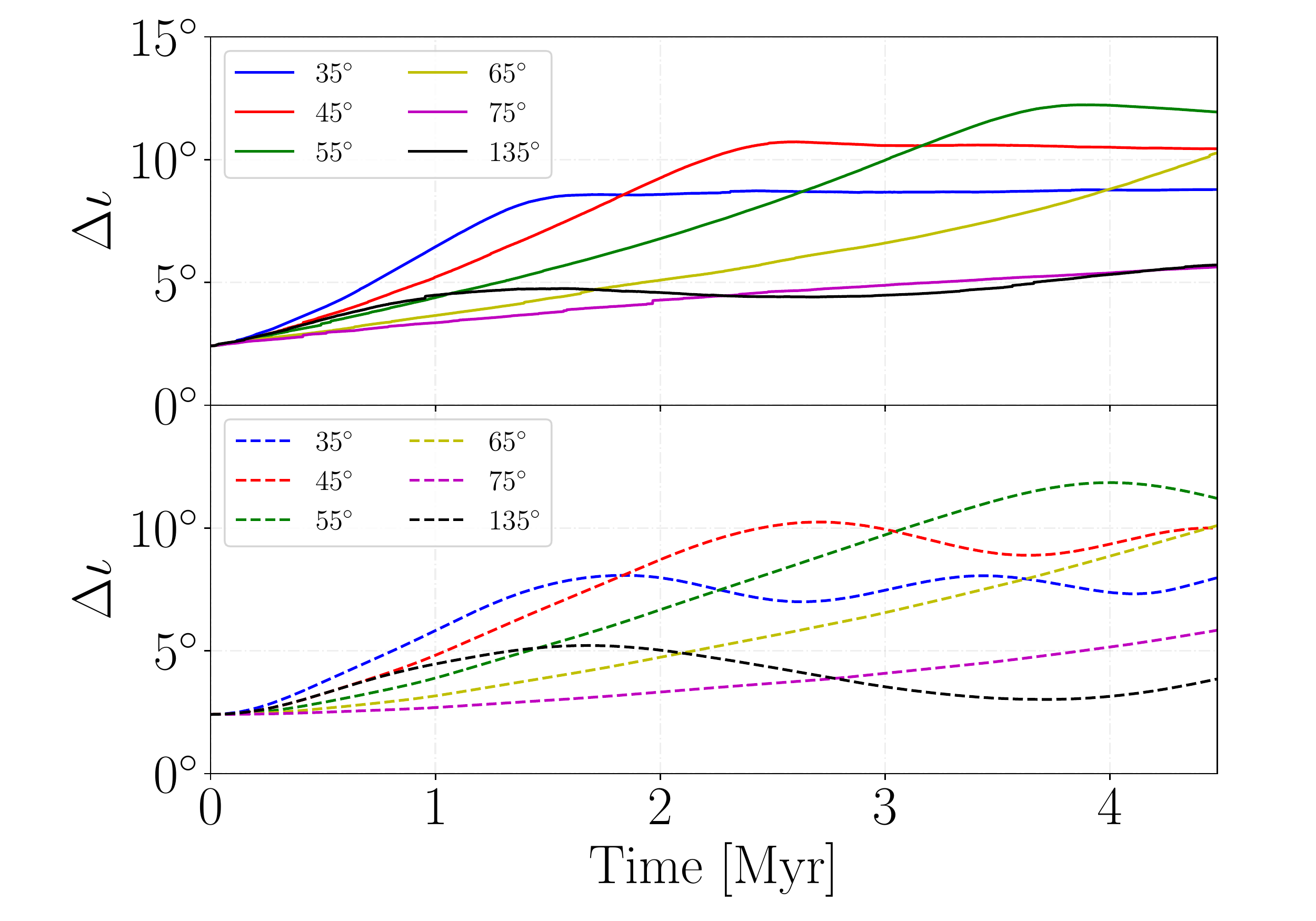}%
\caption{\label{fig:inclination-dependence} 
Inclination evolution of the IMBH using $\varphi$\textsc{GPU} (solid curves) and \textsc{N-ring} (dashed curves) from different initial inclinations. $\iota = \cos^{-1}(\hat{\boldsymbol{L}}_\mathrm{disk} \cdot \hat{\boldsymbol{L}}_\mathrm{IMBH})$, where $\hat{\boldsymbol{L}}_\mathrm{disk}$ is the direction of the total angular momentum of disk stars.
The bottom panel is similar to that of Figure \ref{fig:inclination-t}, but showing the evolution of the warp or thickness of the disk (Eq.~\ref{eq:Di}). Colors refer to different initial inclinations of the IMBH. Top and bottom subpanels of the bottom panel show the results of $\varphi$\textsc{GPU} and \textsc{N-ring}, respectively. See Figures~\ref{fig:inclination-gamma1.75} and \ref{fig:inclination-t-long} for longer timescales and different surface density profiles.}
\end{figure}

Let us now compare the results of models with different initial inclinations. Figure \ref{fig:inclination-dependence} shows the evolution of the IMBH inclination in the top panel with respect to the disk ($\iota = \cos^{-1}(\hat{\boldsymbol{L}}_\mathrm{disk} \cdot \hat{\boldsymbol{L}}_\mathrm{IMBH})$) for $6$ different initial inclinations, $35^\circ$, $45^\circ$, $55^\circ$, $65^\circ$, $75^\circ$, and $135^\circ$, from bottom to top. The bottom panel shows the evolution of $\Delta \iota$ (Eq.~\ref{eq:Di}). Solid and dashed lines show simulations with $\varphi$\textsc{GPU} and \textsc{N-ring}, respectively, started from the same relaxed disk initial conditions (see Section~\ref{sec:initial}). For $\Delta \iota$ this is shown in separate subpanels for clarity, as the curves are intersecting. 

The figure shows that the $\varphi$\textsc{GPU} and \textsc{N-ring} results are in qualitative agreement. The alignment is somewhat more rapid in $\varphi$\textsc{GPU} by a factor $\sim 10\%$ ($\iota_0 = 35^\circ$), $\sim 5\%$ ($\iota_0 = 45^\circ$), and $\sim 4\%$ ($\iota_0 = 55^\circ$) than in the \textsc{N-ring} results. Regardless of the value $\iota_0$, the discrepancy is more significant once the alignment angle is less than $\iota\lesssim 30^\circ$. The difference is well-explained by the differences in the approximations; \textsc{N-ring} neglects two-body relaxation and SRR by assuming that the eccentricity and semimajor axes of all objects are conserved. As shown above by Eq.~\eqref{eq:talign}, energy exchange by close two-body encounters results in Chandrasekhar dynamical friction, which leads to an alignment timescale of $43$ Myr at $\iota=45^{\circ}$, $9.4$ Myr at $30^{\circ}$ and $1.9$ Myr at $20^{\circ}$. The total time needed for alignment is determined by the evolution at large $\iota$ if $\iota_0>30^{\circ}$, where Chandrasekhar dynamical friction is insignificant. Note further that Figure~\ref{fig:a-e-t} has shown that the eccentricity of the IMBH changes significantly over a Myr, which is a manifestation of scalar resonant relaxation. While this is also neglected in \textsc{N-ring}, VRR is relatively insensitive to eccentricity for $e<0.7$ as shown in Figure 7 of \citet{Kocsis2015}. Thus, we conclude that the alignment is predominantly driven by VRR if $\iota_0>30^{\circ}$, and in this case the total alignment time is well-modeled by \textsc{N-ring}.

The counter-rotating simulations do not lead to systematic deviations between the two methods, showing that VRR is dominant over scalar resonant relaxation and two-body relaxation in this case. Here, the IMBH's inclination angle and the disk exhibit small-amplitude periodic oscillations. In this case, the conservation of total angular momentum vector inhibits the IMBH's orbital flip. For counter-rotating orbits, two-body interactions between disk stars and the IMBH is extremely weak to drive angular momentum exchange between the IMBH and the disk. We ran the \textsc{N-ring} simulations for an extended timescale of 150 Myr for $\iota_0=105^{\circ}$ and $135^{\circ}$, and found small-amplitude oscillations throughout the evolution.

The similarity of the results of the two numerical methods at large inclinations $\iota_0 \gtrsim 30^\circ$ in Figure~\ref{fig:inclination-dependence}, shows that the IMBH's alignment is mainly driven by VRR. \citet{Rauch1996} named this process resonant dynamical friction (RDF). Next, we explore this process by running \textsc{N-ring} simulations on longer timescales, different IMBH masses, and surface density profiles.

\subsection{Dependence on the IMBH mass}
\label{sec:IMBHmass}
\begin{figure}
\centering 
  \includegraphics[width=0.5\textwidth]{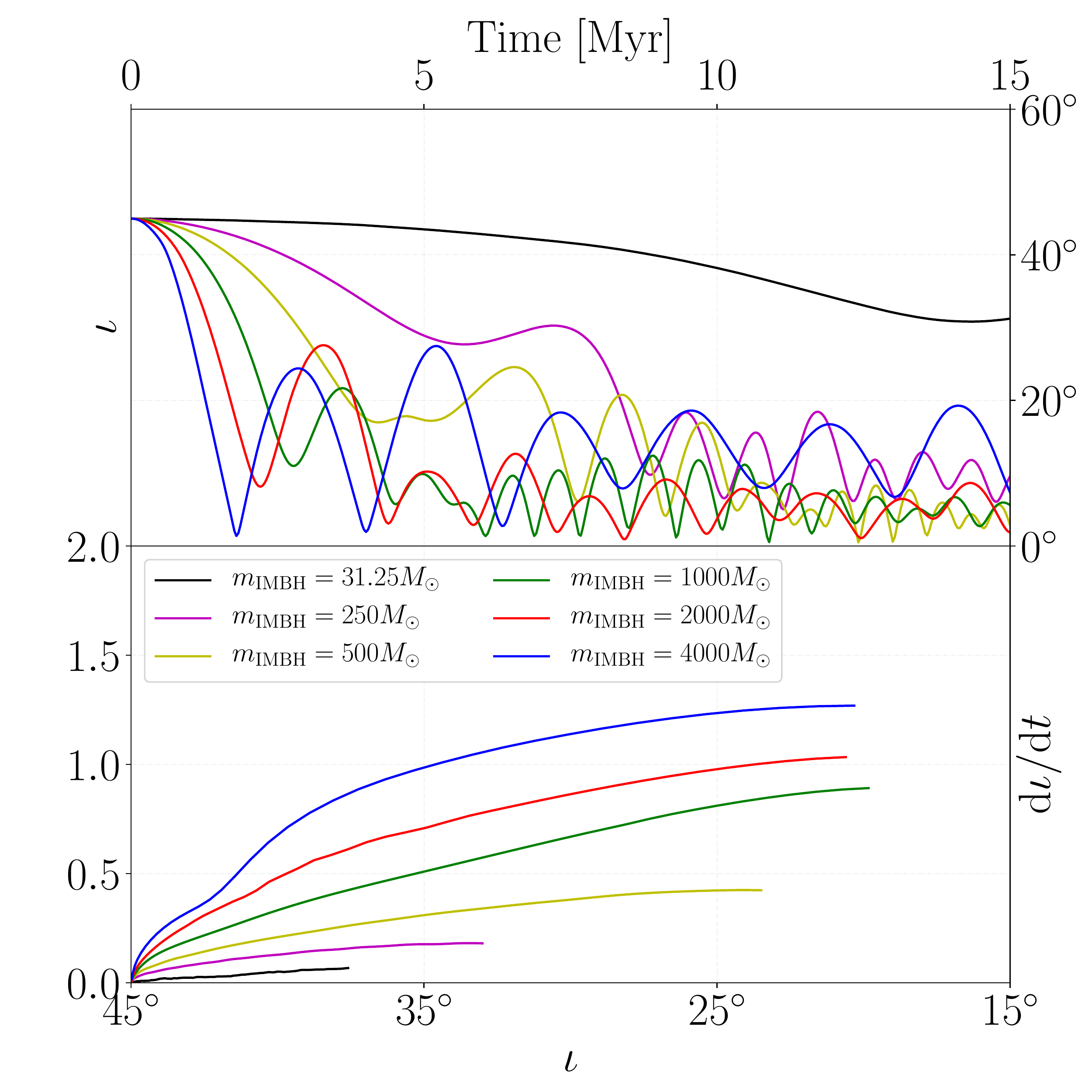}\\
  \includegraphics[width=0.5\textwidth]{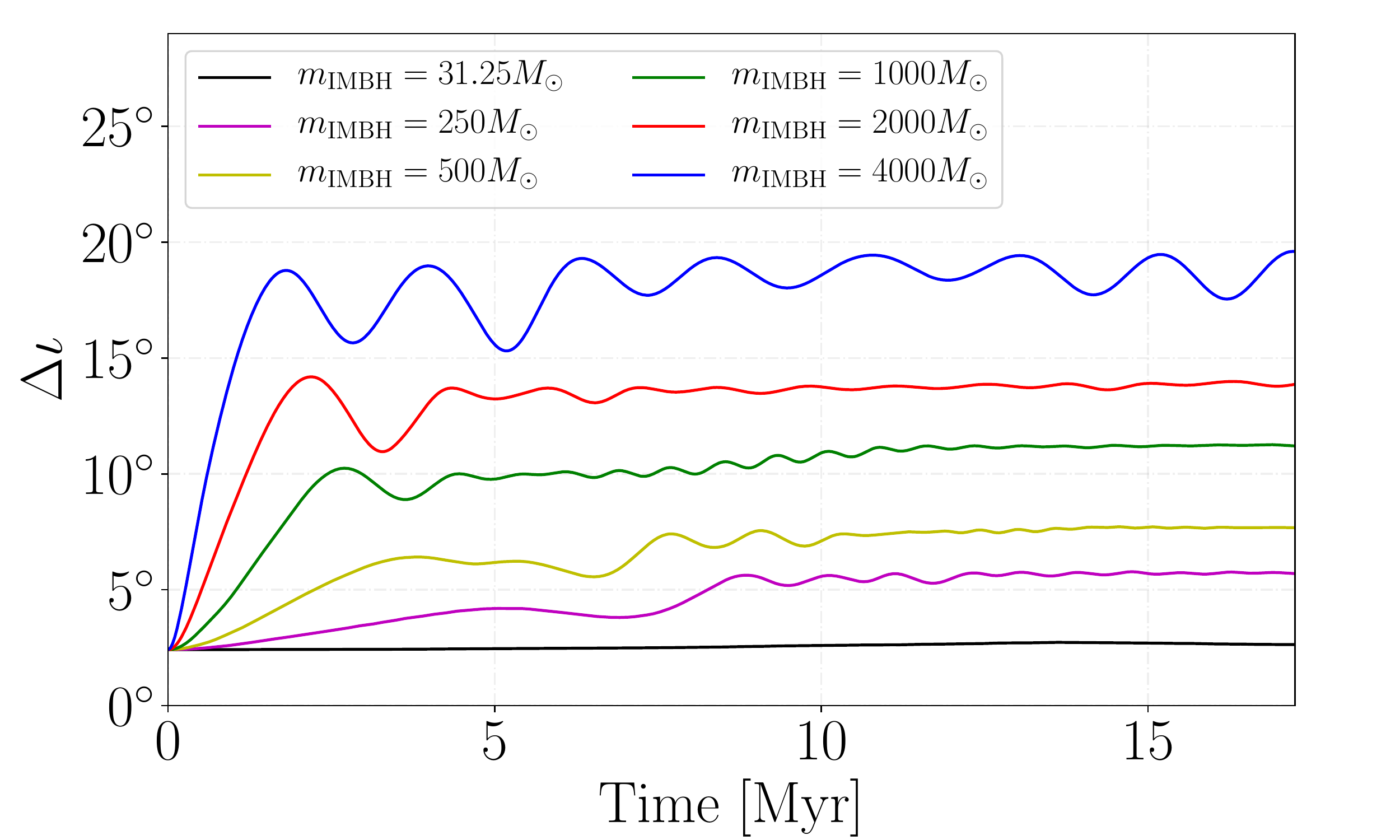}
  \caption{\label{fig:mass-dependence}
  \textit{Top:} The evolution of the orbital inclination of an IMBH with a mass of $31.25$, $250$, $500$, $1000$, $2000$, or $4000 \Msun$, respectively, due to a stellar disk starting from $\iota_0 = 45^\circ$ initial inclination in \textsc{N-ring}. The top subpanel is similar to the top panel of Figure \ref{fig:inclination-dependence}, i.e. the inclination of the IMBH as a function of time. The bottom subpanel shows the the rate of change of the inclination as a function of inclination in the first decreasing phases, prior to the first local minima in the top panel. \textit{Bottom:} Similar to the bottom panel of Figure \ref{fig:inclination-dependence} showing the warp or thickness of the disk as a function of time for different IMBH masses (Eq.~\ref{eq:Di}). 
  }
\end{figure}

\begin{figure}
\centering 
  \includegraphics[width=0.5\textwidth]{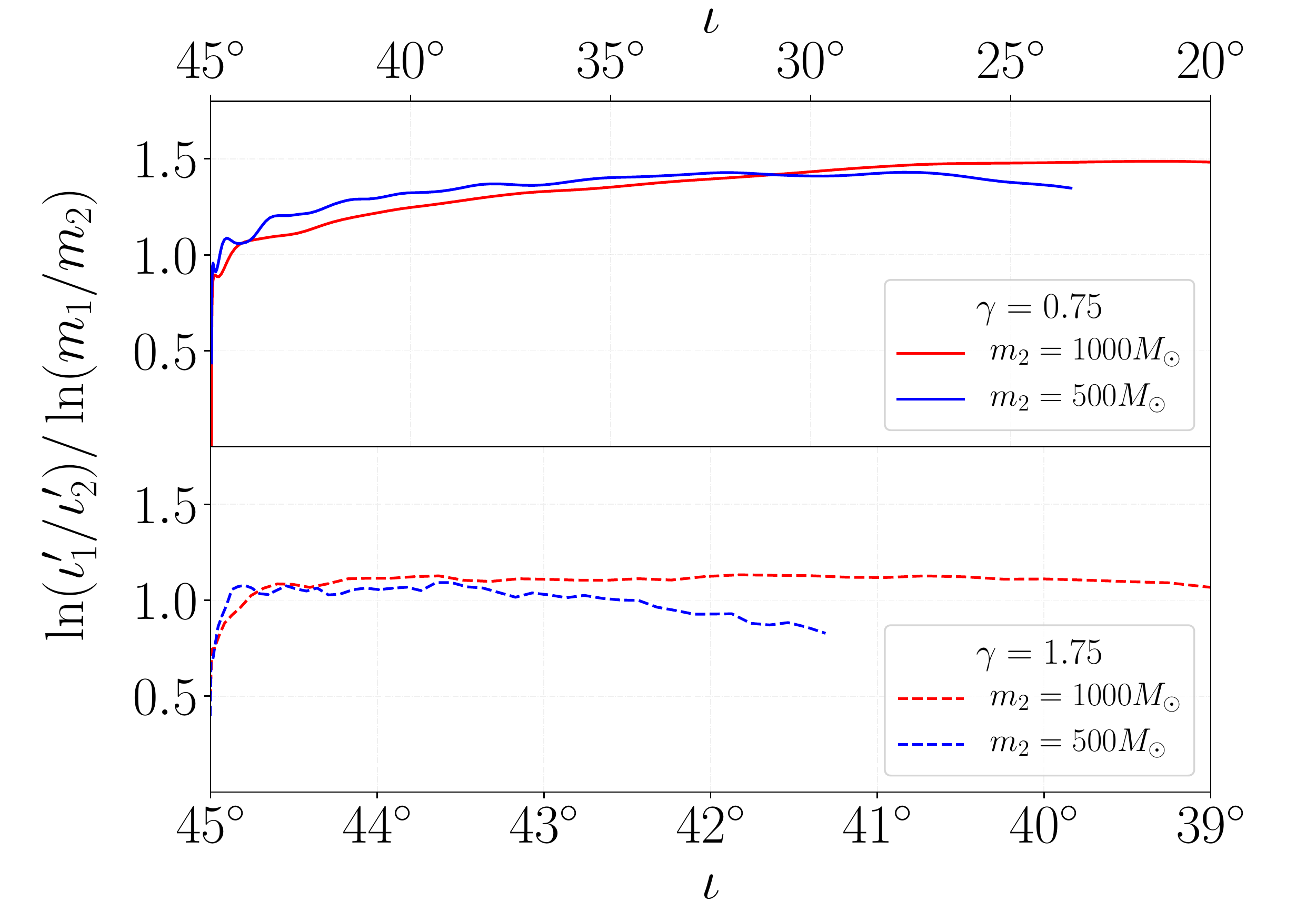}%
  \caption{\label{fig:logiperlogm} The $\alpha$ exponent of the mass dependence of the alignment rate, $\rm{d}\iota/\rm{d}t \propto m^\alpha$, measured as a function of  inclination in a disk with surface density exponent $\gamma=0.75$ (top) and $1.75$ (bottom). We compare simulations with IMBH mass $m_1=250\,\Msun$ versus $m_2 = 500\,\Msun$ (blue) or $m_2 = 1000 \Msun$ (red) and plot $\ln(\iota'_1/\iota'_2) / \ln(m_1/m_2)$ where $\iota'_1 = \mathrm{d}\iota_1 / \mathrm{d}t $. The value of $\iota_1$ is the inclination of IMBH with mass $m_1$ and similarly for $\iota'_2$. All simulations were initialized with $\iota_1=\iota_2=45^\circ$. In all cases, the mass dependence is close to linear, $\alpha=1$, initially.
  }
\end{figure}

Let us now compare the alignment times for a fixed initial IMBH inclination ($\iota_0 = 45^\circ$) with different IMBH masses, $m_\mathrm{IMBH} = 31.25$, $250$, $500$, $1000$, $2000$, $4000 \Msun$. The top panel in Figure \ref{fig:mass-dependence} shows the IMBHs' inclination with respect to the disk as a function of time (top subpanel) and $\mathrm{d}\iota / \mathrm{d}t$ as a function of inclination (bottom subpanel), while the bottom panel shows the warp/thickness of the disk. We find that, for an IMBH mass of $250 \leq m_{\rm IMBH} \leq 1000 \Msun$, the initial rate of alignment is approximately proportional to 
\begin{equation} \label{eq:mass_dep}
    \left.\frac{\rm{d}\iota}{\rm{d}t}\right|_{\rm RDF} \propto m^{\alpha}_{\rm IMBH}
\end{equation}

We measure $\alpha$ by comparing pairs of simulations as $\alpha = \ln[(\rm{d}\iota_1/\rm{d}t)/(\rm{d}\iota_2/\rm{d}t)]/\ln(m_1/m_2)$ for masses (250, 500, or $1000\,\Msun$) starting from the same $\iota=45^\circ$. Figure \ref{fig:logiperlogm} shows that $\alpha=1\pm 0.1$ initially for $44^{\circ}\leq\iota\leq45^{\circ}$ for both $\gamma=0.75$ and for $\gamma=1.75$. At later times for $25^\circ\leq \iota<44^\circ$, $\alpha=1.3\pm 0.2$ for $\gamma=0.75$, but it is less than $1$ for $\gamma=1.75$ for $41^\circ\leq \iota<44^\circ$, as the inclination exhibits a hang-up (see Section~\ref{sec:gamma=1.75} below). Note that $\alpha=1$ matches the value for CDF. For larger IMBH masses, the mass dependence of alignment rate is shallower, which is to be expected since in this case the local disk mass is comparable to or smaller than the IMBH mass, and angular momentum conservation inhibits the rapid relaxation process. 

We find that the IMBH ultimately aligns with the stellar disk for $m_{\rm IMBH}\geq 250\Msun$ for our standard model with $(m_{\rm d}, \gamma, N, \iota_0)=(8191\Msun, 0.75, 8191, 45^{\circ})$ and also for our alternative model with $(m_{\rm d}, \gamma, N, \iota_0)=(8191\Msun, 1.75, 8191, 35^{\circ})$. However, there is no alignment for $m_{\rm IMBH}= 31.25\Msun$ for the former model. In this case, the system exhibits quasiperiodic oscillations between $30^{\circ}\leq \iota \leq 45^{\circ}$ throughout the simulation for 175 Myr \citep[see also][for a discussion of normal mode oscillations of a thin nuclear stellar disk]{Kocsis2011}.

\subsection{Dependence on the disk mass, radius, SMBH mass} \label{sec:tunit}
Due to the scale-free nature of the \textsc{N-ring} simulations, the simulation time in code units is converted to physical time by scaling with
\begin{equation}\label{eq:tunit}
    t_{\rm unit} = \frac{r_{\rm unit}\sqrt{G m_{\rm SMBH} r_{\rm unit}}}{G m_{\rm unit}} = \frac{m_{\rm SMBH}}{m_{\rm unit}} \frac{t_{\rm orb}(r_{\rm unit})}{2\pi}
\end{equation}
where $r_{\rm unit}$ and $m_{\rm unit}$ are the distance and mass units adopted in the simulation. Here, $t_{\rm orb}(r)=2\pi (G m_{\rm SMBH})^{-1/2}r^{3/2}$ is the orbital time. Note that the \textsc{N-ring} simulation in code units is independent of $m_{\rm SMBH}$; this quantity enters the results only when converting the code units to physical units using Eq.~\eqref{eq:tunit}. Indeed, \textsc{N-ring} follows orbit-averaged interactions among stellar objects; the influence of the SMBH is accounted for by the orbit averaging. Eq.~\eqref{eq:tunit} shows that, if the mass of the IMBH and the disk are both increased by the same factor $\kappa$, keeping the SMBH mass and all other parameters fixed, this is equivalent to changing $m_{\rm unit}$ to $\kappa m_{\rm unit}$ and the evolution is identical if the time is scaled by the factor $\kappa^{-1}$.

However, we have shown in Section \ref{sec:IMBHmass} that the alignment rate scales initially approximately with $m_{\rm IMBH}^\alpha$ where $\alpha\approx 1$ initially, and later by a somewhat different value for a range of IMBH and disk masses with $m_{\rm IMBH}\leq m_{\rm d,loc}$. Based on dimensional analysis,\footnote{Indeed, denoting the dimensionless code units with overline: $\bar{m}_{\rm IMBH} = m_{\rm IMBH}/m_{\rm unit}$, $\bar{m}_{\rm d,loc} = m_{\rm d,loc} / m_{\rm unit}$, $\bar{t} = t / t_{\rm unit}$, $\rm{d}\iota / \rm{d}\bar{t} = t_{\rm unit} \rm{d}\iota / \rm{d}t$ and given that the evolution in code units is independent of $m_{\rm SMBH}$, assuming that $\rm{d}\iota/\rm{d}\overline{t} \propto \overline{m}_{\rm IMBH}^{\alpha} \overline{m}_{\rm d,loc}^{\beta} $ for $\alpha$ and $\beta$ real numbers, implies that $\rm{d}\iota/\rm{d}t \propto
t_{\rm unit}^{-1} (m_{\rm IMBH}/m_{\rm unit})^{\alpha}(m_{\rm d,loc}/m_{\rm unit})^{\beta}$. Substituting $t_{\rm unit}$ from Eq.~\eqref{eq:tunit} shows that $\beta=1-\alpha$ to ensure that $\rm{d}\iota/\rm{d}t$ is independent of the assumed units.} combining with Eq.~\eqref{eq:tunit} implies that if the RDF rate of alignment scales with the masses as $\rm{d}\iota/\rm{d}t \propto m_{\rm IMBH}^{\alpha} m_{\rm d,loc}^{\beta} m_{\rm SMBH}^{\delta}$ then $\beta=1-\alpha$ and $\delta = 1$. For $\alpha\approx 1$ this is independent of $m_{\rm d,loc}$. Note, however, that here $\iota$ is the relative angle between the IMBH and the disk, and the rate of reorientation of the IMBH only (with respect to an inertial frame) may be different if the disk angular momentum is comparable to or less than the IMBH's (see Section \ref{sec:analyticRDF}). 

Furthermore, Eq.~\eqref{eq:tunit} shows that, if the semimajor axes of the IMBH and all disk stars are scaled by a factor $\kappa$, this results in an identical evolution if the time unit is increased according to the change in the orbital time, i.e., by a factor of $\kappa^{3/2}$.
\begin{figure*}
\centering 
\includegraphics[width=1.0\textwidth]{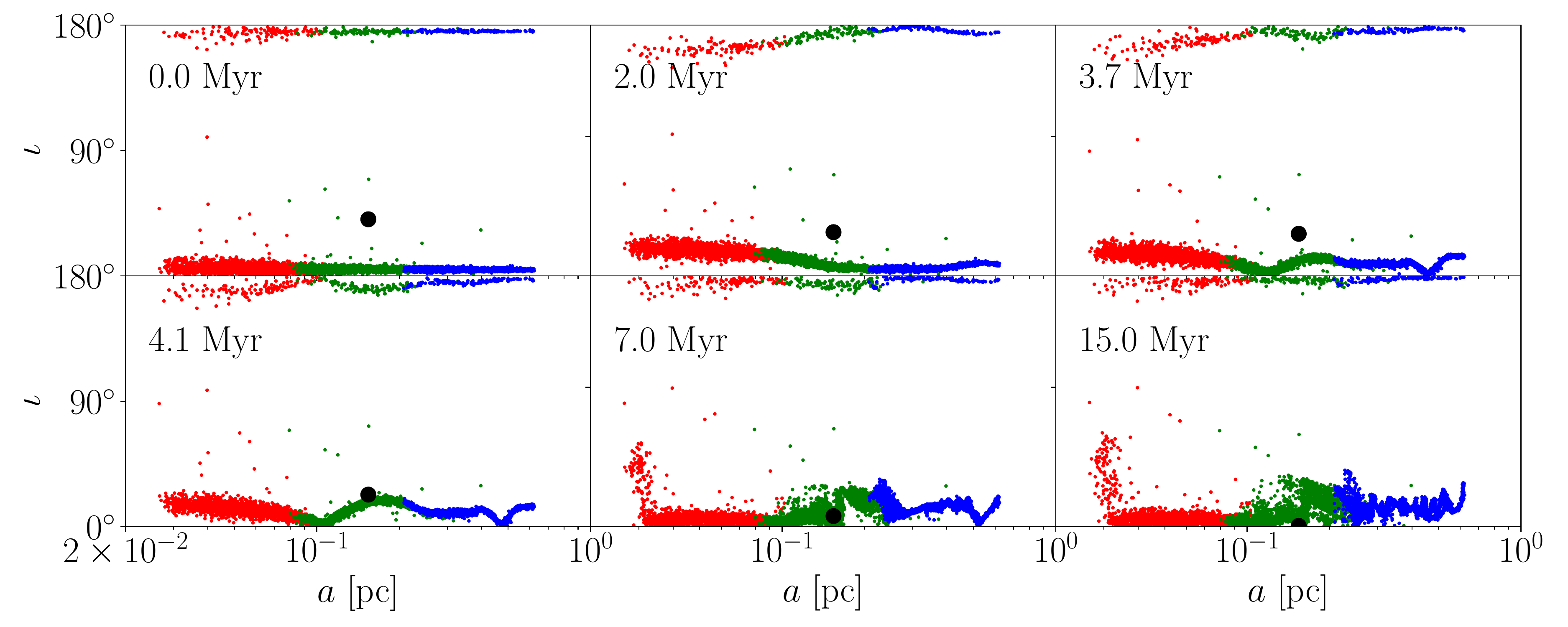}
\caption{\label{fig:snapshots_Nring_175_i45} Same as Figure~\ref{fig:snapshots} but for a surface density profile that scales with $r^{-1.75}$ instead of $r^{-0.75}$ showing the scatter plot of semimajor axes and orbital inclinations of the disk and the IMBH for 6 representative snapshots in the 6 panels as labeled. The initial condition (top left panel) has $\iota_0=45^\circ$. There are more stars in the inner region (red points) and the inner disk gets warped less than in Figure~\ref{fig:snapshots}. }
\end{figure*}
\begin{figure*}
\centering 
\includegraphics[width=1.0\textwidth]{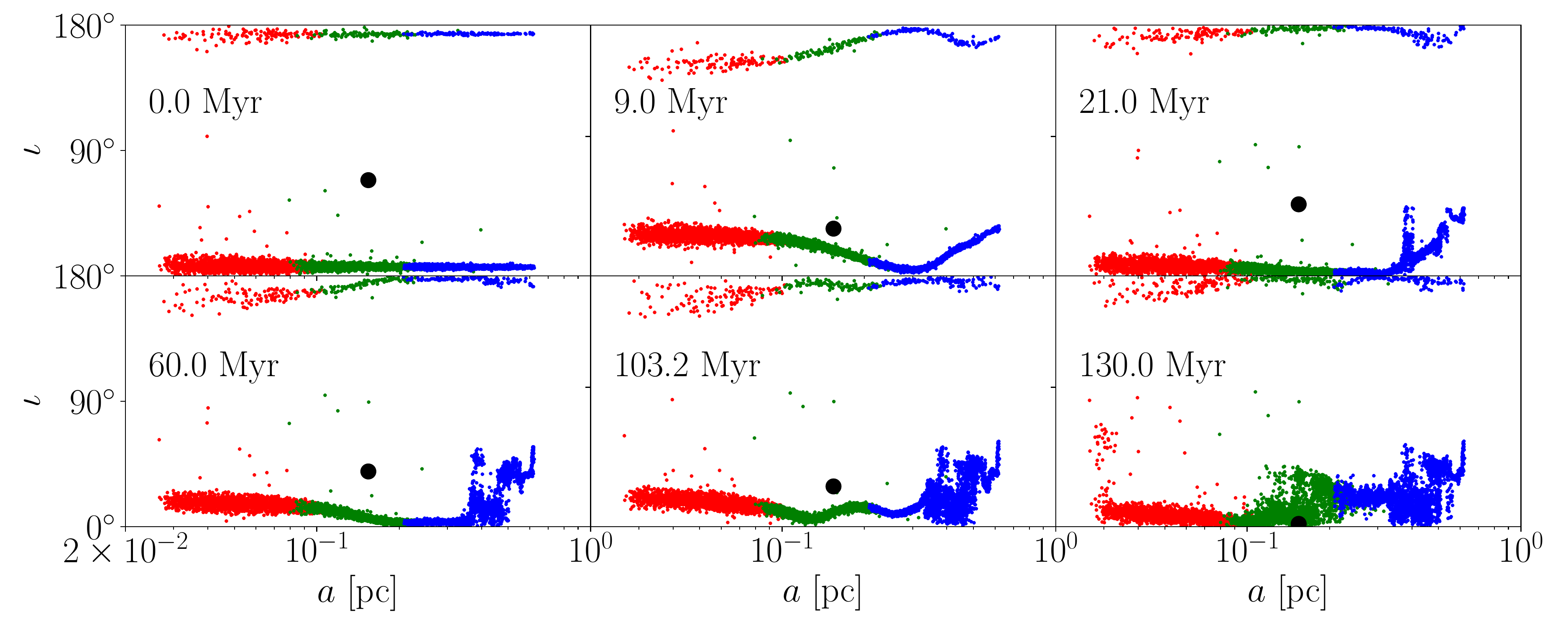}
\caption{\label{fig:snapshots_Nring_175_i75} Same as Figure~\ref{fig:snapshots_Nring_175_i45} but for initial IMBH orbital inclination of $\iota_0=75^\circ$. The alignment time is much longer, as the IMBH stalls at around $45^{\circ}$ for an extended period of time until a discontinuity develops in the outer disk and propagates inward to the IMBH radius.}
\end{figure*}

\subsection{Dependence on the number of stars}
To examine how the results depend on the total number of disk stars $N$, we compare simulations with approximately fixed total disk mass and IMBH mass, i.e.~$(N, m_\mathrm{disk}, m_\mathrm{IMBH}) = (4095, 8190\,\Msun, 2000\,\Msun)$ with $(8191, 8191\,\Msun, 2000\,\Msun)$ and $(2047, 8188\,\Msun$, $4000\,\Msun)$ with $(8191, 8191\,\Msun, 4000\,\Msun)$, and find that the respective discrepancies in the alignment time are less than $10\%$ and $16\%$. These simulations assume $\gamma=0.75$ and $\iota_0=45^{\circ}$. We conclude that resonant dynamical friction in many cases does not depend on the mass of individual disk stars and their total number for fixed stellar surface density profile. 

However, we found that the number of particles $N$ significantly affects the long-term evolution for surface density profile $\gamma=1.75$, high initial IMBH inclination, and IMBH masses less than the local disk mass. The time duration of the orbital inclination hang-up phase is increased for higher $N$ and the oscillation amplitude is decreased in this phase; see further discussion in Sec.~\ref{sec:gamma=1.75}.

\subsection{Dependence on the disk surface density profile, $\gamma=1.75$}\label{sec:gamma=1.75}
Up to this section, we mostly used stellar disk surface density profile $\rho\propto r^{-\gamma}$ with $\gamma = 0.75$ in Eq.~\eqref{eq:rho}. Now
using \textsc{N-ring},
we present the time evolution for a stellar disk with a steeper radial density profile with $\gamma = 1.75$, consistent with that in the Galactic Center \citep{Bartko2009,Lu2009,Yelda2014}. We keep all other parameters of the surface density profile (Eq.~\ref{eq:rho}) and the IMBH the same as before. In this case, the local disk mass is $1.5\times$ smaller at the IMBH semimajor axis, and the mass distribution is approximately uniform on a log scale $m_{\rm d, loc} \propto r^{0.25}$, while for $\gamma=0.75$ most of the mass was at the outside $m_{\rm d, loc} \propto r^{1.25}$.

\begin{figure}
\centering 
\includegraphics[width=0.5\textwidth]{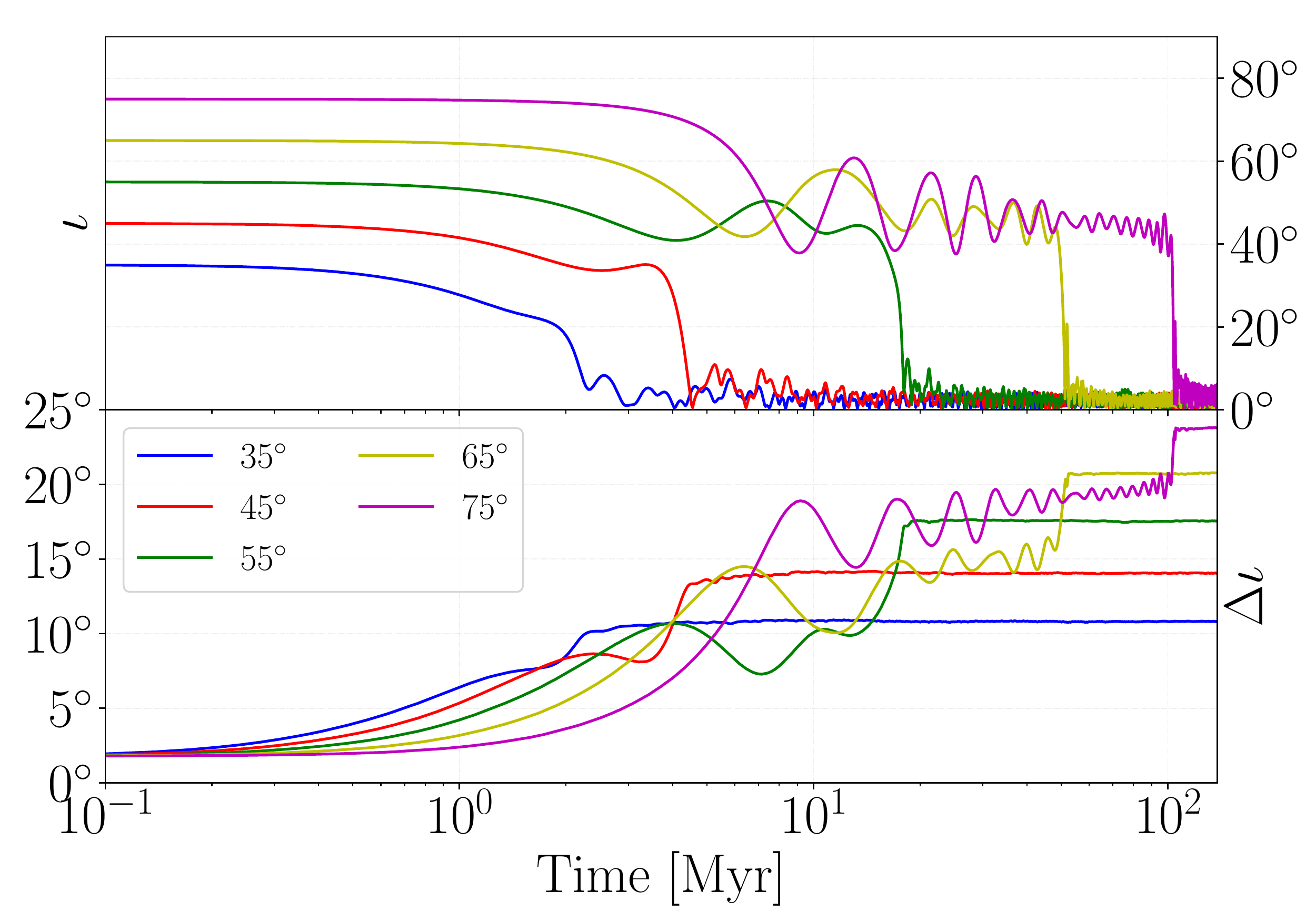}%
\caption{\label{fig:inclination-gamma1.75} Similar to Figure \ref{fig:inclination-dependence} showing the inclination evolution of the IMBH (top panel), as well as the evolution of disk warp and thickness (bottom panel), using \textsc{N-ring} but with a disk model with surface density profile $\propto r^{-1.75}$. Here, $m_{\rm IMBH}=1000\Msun$, and the total mass ($8191\Msun$) and number of stars in the disk ($N=8191$) are the same as in Figures~\ref{fig:inclination-dependence}. The evolution consists of three phases: (i) initial reorientation, (ii) oscillations around an intermediate orbital inclination of $\iota\sim 45^\circ$, and (iii) final plunge. Phase (ii) prolongs the total alignment time significantly for $\iota_0\geq 55^\circ$ (see Figures~\ref{fig:inclination-dependence} and \ref{fig:diff-disk-particles}).}
\end{figure}

We initialize the IMBH ($m_\mathrm{IMBH} = 1000 \Msun$) with $\iota_0 = 35^{\circ}$, $45^\circ$, $55^\circ$, $65^\circ$ and $75^\circ$, respectively, and follow the evolution of the system. Figures \ref{fig:snapshots_Nring_175_i45} and \ref{fig:snapshots_Nring_175_i75} show snapshots of the evolution of orbital inclinations vs. semimajor axes for $45^\circ$ and $75^\circ$, respectively. In contrast to the case with $\gamma=0.75$ where the inner disk gets warped rapidly, Figure~\ref{fig:snapshots}, the inner disk remains flat for $\gamma=1.75$ and $\iota_0\geq 55^\circ$ and the outer disk develops the warp starting from the outer edge. The warp propagates inward in time, at a rate that is smaller for larger $\iota_0$. If $\iota_0\geq 55^{\circ}$, the IMBH orbital inclination exhibits a hang-up until the disk warp reaches close to the IMBH apoapsis, after which the IMBH quickly plunges into the disk. For highly inclined initial conditions, the orbital hang-up may be quite prolonged; see the yellow and purple curves in the top panel of Figure \ref{fig:inclination-gamma1.75}.

Figure \ref{fig:inclination-gamma1.75} shows the inclination of the IMBH (top) and the evolution of the disk warp/thickness (bottom) as a function of time for the 5 different cases. In comparison to the $\gamma = 0.75$ case (see Figure \ref{fig:inclination-dependence} and \ref{fig:inclination-t-long}), here the local disk mass is $1.5\times$ smaller, and the alignment occurs more slowly by $\sim 1.2 \times$ ($\iota_0 = 35^\circ$) and $\sim 1.6 \times$ ($\iota_0 = 45^\circ$). For larger $\iota_0$ the alignment time is much longer due to the orbital hang-up by $\sim 4.3 \times$ ($\iota_0 = 55^\circ$), $\sim 8.5 \times$ ($\iota_0 = 65^\circ$), and $\sim 12.4 \times$ ($\iota_0 = 75^\circ$), respectively. In this case, the IMBH does not follow the same exponential decay as the empirical fit for $\gamma = 0.75$, Eq.~\eqref{eq:inc}. For simulations with $\iota_0 = 55^\circ$, $65^\circ$, $75^\circ$, the IMBH inclination first decreases to an intermediate value of $\sim 45^\circ$ ) where the IMBH oscillates with a decreasing amplitude until structural changes in the disk allow the IMBH's direct and rapid infall. 

Note that Chandrasekhar's dynamical friction has been neglected here, but that timescale is even longer than the alignment time found here (see the solid line and $\times$ symbols in Figure~\ref{fig:empirical-fit}). CDF may possibly affect the evolution for $\iota_0 \gtrsim 55^\circ$ by slowly pushing the orbital inclination past the intermediate equilibrium value, thereby possibly shortening the orbital hang-up phase. A detailed study of the orbital hang-up phase requires direct N-body simulations over extended time periods, which are beyond the scope of this paper.

\begin{figure}
\centering 
  \includegraphics[width=0.5\textwidth]{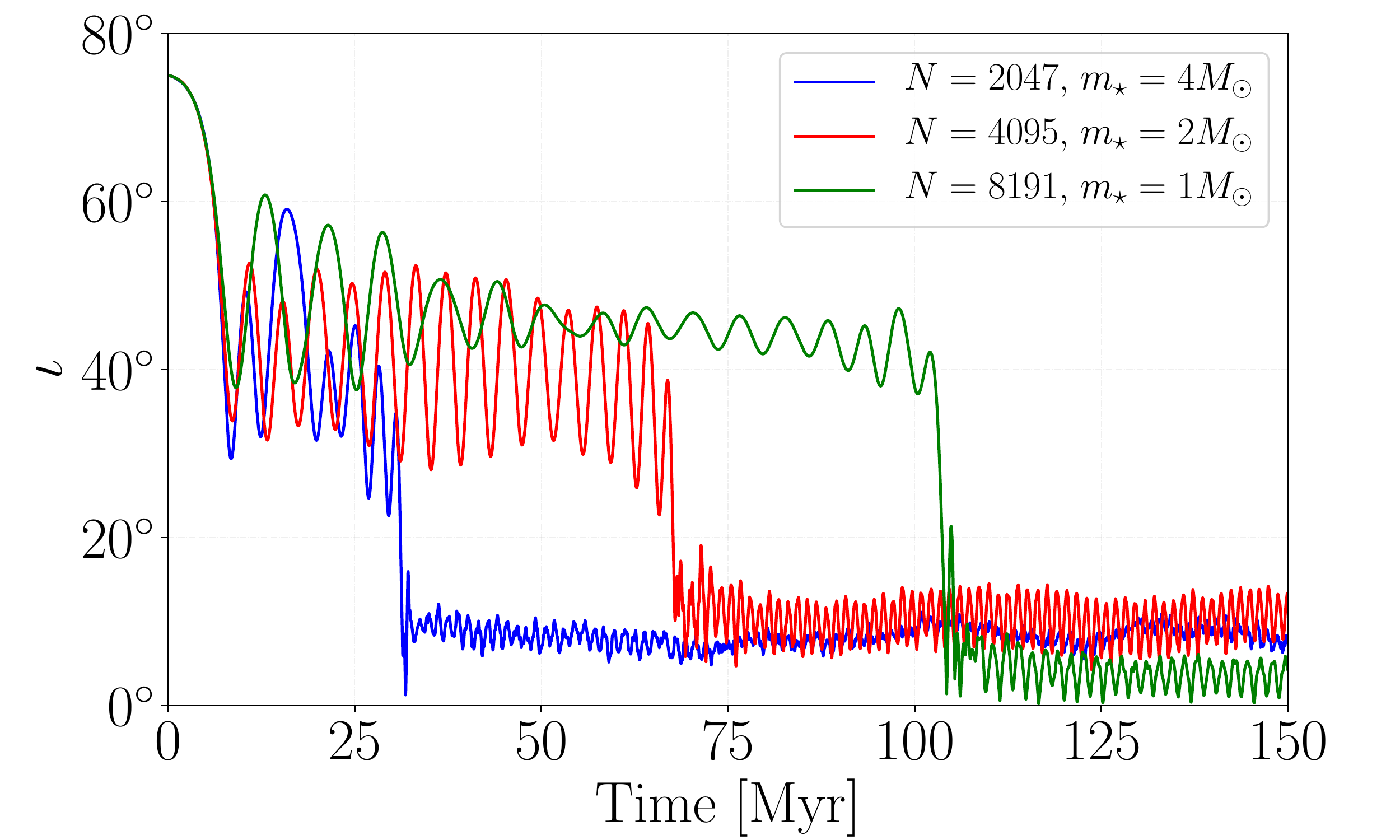}%
\caption{\label{fig:diff-disk-particles} Similar to the top panel in Figure \ref{fig:inclination-gamma1.75}: the orbital inclination of the IMBH for $\iota_0=75^\circ$ as a function of time relative to a stellar disk containing different numbers of disk stars $N = 2047$ (blue), $4095$ (red), and $8191$ (green), respectively. Other parameters are the same as in Figure \ref{fig:inclination-gamma1.75} including the mass of the IMBH  ($1000 \Msun$), as well as the total mass and surface density profile of the disk: $8191\Msun$ and $r^{-1.75}$.}
\end{figure}

Figure~\ref{fig:diff-disk-particles} shows how the number of disk stars, $N$, affects the orbital inclination hang-up phase for model ($m_{\rm IMBH}$, $\iota_0$, $\gamma$) = ($1000 \Msun$, $75^{\circ}$, $1.75$) for $N=2047$, $4095$, and $8191$ and fixed total disk mass $m_{\rm d}=8191\Msun$. We find that the oscillation amplitude is increased around the intermediate equilibrium inclination of $45^{\circ}$ and the total time duration of the hang-up is decreased for smaller $N$. It is remarkable that, for systems that exhibit the orbital inclination hang-up phenomenon, the evolution remains sensitive to $N$ even in the $N\gg 1000$ limit. Indeed, Figure~\ref{fig:diff-disk-particles} shows that, while the initial rate of reoriention is independent of $N$ for a fixed initial $\iota$, the orbital hang-up phase may be significantly extended in time for very large $N$. We find that this is due to a discontinuity that forms in the inclination distribution in the disk in a narrow range of semimajor axis in the outer region, i.e. near $a=0.38\,\rm pc$ at 21 Myr for $N=8191$ in Figure~\ref{fig:snapshots_Nring_175_i75}. However, since the discontinuity is manifested in only a small number of stars, the evolution remains sensitive to $N$ even for $N=8191$. The final IMBH realignment takes place after the discontinuity propagates to overlapping radii. The discontinuity moves more slowly for a smoother disk, i.e. for larger $N$.

\subsection{Analytic model} \label{sec:analyticRDF}
Following the arguments in Sec.~\ref{sec:tunit}, in the limit of a scale-free disk, and with an IMBH much larger than the mass of individual objects in the disk, we expect a scaling as\footnote{Here, $m_{\rm d}$ may depend on the total disk mass, the local disk mass (Eq.~\ref{eq:mloc}), or the characteristic disk mass (Eq.~\ref{eq:mdchar}). We leave a more accurate determination to future work.}
\begin{equation}\label{eq:RDF_scaling}
    \left.\frac{\rm{d}\iota}{\rm{d} t}\right|_{\rm RDF} = f\left(\iota,\iota_0,\frac{m_{\rm IMBH}}{m_{\rm d}}\right) \frac{m_{\rm d}^{1-\alpha}m_{\rm IMBH}^{\alpha} \,t_{\rm orb}^{-1}}{m_{\rm SMBH}} 
\end{equation}
where $\alpha\approx 1$ initially as measured above (and it may be somewhat larger or smaller at later times) and $f$ may depend on the IMBH inclination with respect to the actual state of the disk, which in turn may be expected to depend mostly on the initial value $\iota_0$, the ratio of IMBH mass to the local disk mass and the IMBH mass to the individual object mass in the disk. Note that, as long as $\iota \approx \iota_0$, the disk has not had time to change significantly and $f$ is expected to depend mostly only on $\iota_0$. We determine this function by fitting to simulations next. We derive this function for the model with $\gamma=0.75$ since this model is free from discontinuities and does not exhibit the orbital hang-up phenomenon, but we find that the resulting formula is a reasonable approximation to describe the initial rate of reorientation for the $\gamma=1.75$ model as well.

\begin{figure}
\centering 
  \includegraphics[width=0.5\textwidth]{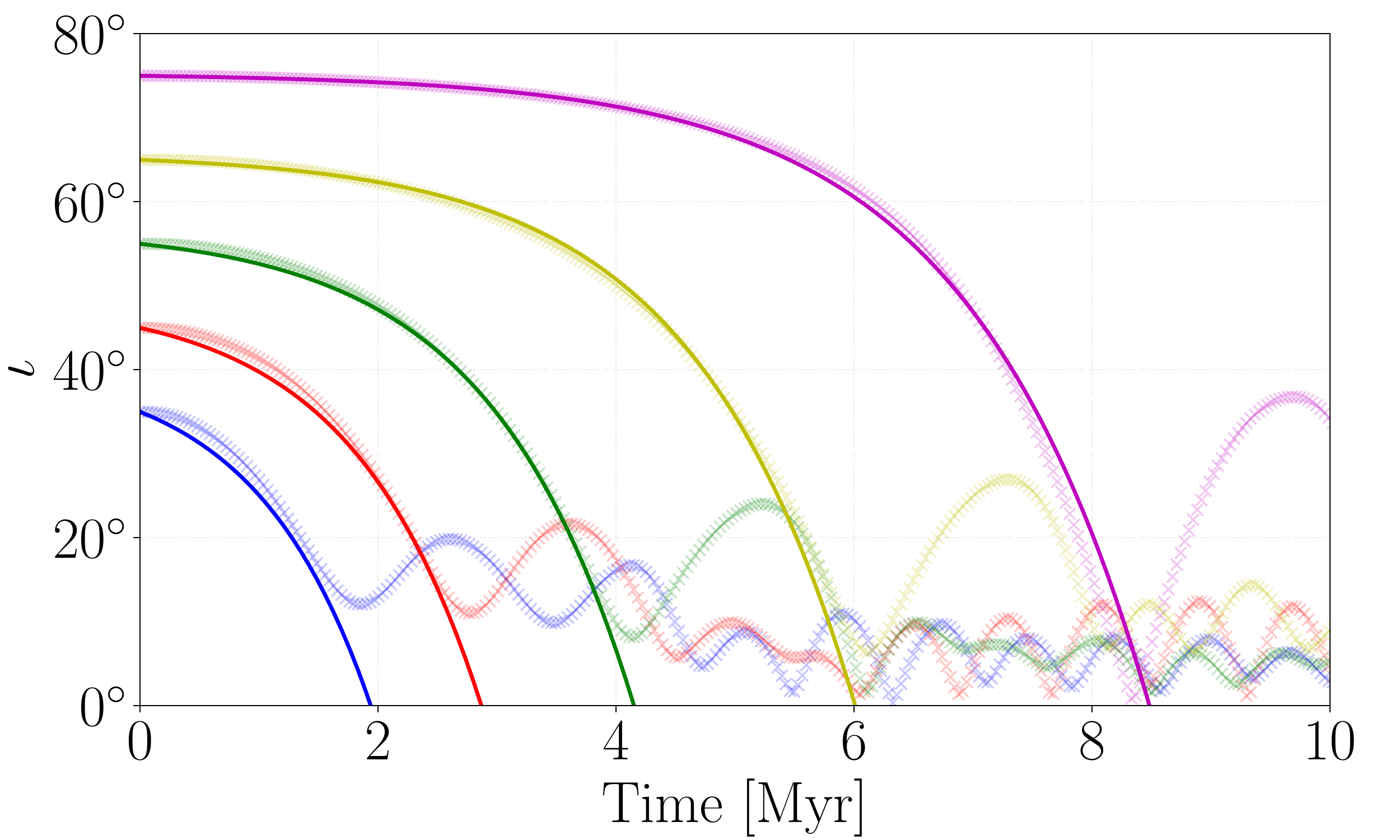}%
\caption{ \label{fig:inclination-t-long} Similar to Figure \ref{fig:inclination-dependence} but showing the longer time evolution. Faded data points show the results of \textsc{N-ring}, and solid lines are the empirical fit, Eq.~\eqref{eq:inc}. The inclination changes exponentially in time.}
\end{figure}
Figure \ref{fig:inclination-t-long} shows the results of \textsc{N-ring} for the evolution similar to the top panel of Figure~\ref{fig:inclination-t} for the same surface density profile Eq.~\eqref{eq:rho} with $\gamma=0.75$, but on longer timescales, and solid lines show empirical fits to the evolution. The inclination angle decays at an exponentially growing rate in the simulations for $\iota_0 \gtrsim 15^\circ$, which may be fit for $35^\circ\leq \iota_0 \leq 75^\circ$ initial inclination by
\begin{equation}\label{eq:inc}
\iota(t) =  \left(\iota_{\min} + \frac{|\iota'_{\max}|}{\lambda}\right)(1-e^{\lambda t})+\iota_0 e^{\lambda t}\,,
\end{equation}
where the parameters depend on $\iota_0$ as
\begin{align}
    \lambda &= 8.2 e^{-0.636\iota_0} \,\frac{m_{\rm IMBH}}{m_{\rm SMBH}} t_{\rm orb}^{-1}  =  1.51 e^{-0.636\iota_0}\,\mathrm{Myr^{-1}}  \,, \\
    \iota'_{\max}&= \, -3.2 \frac{m_{\rm IMBH}}{m_{\rm SMBH}} t_{\rm orb}^{-1} = -0.6\,\mathrm{Myr^{-1}} \,,\\
    \iota_{\min} &= -0.122 + 0.4\iota_0 \,
\end{align}

These parameters have the following physical interpretation:
$\iota'_{\max}$ is the rate of alignment when the IMBH inclination reaches the edge of the warped disk, which is found to be approximately independent of the initial condition $\iota_0$; $\iota_{\min}$ is the inclination when the IMBH reaches the edge of the warped disk, which increases with $\iota_0$ to roughly the saturation value of $\Delta \iota$ (Figure~\ref{fig:inclination-t});and
$\lambda$ is related to the change of the alignment rate $(\iota'_0-\iota'_{\max})/(\iota_0 - \iota_{\min})$.
Note that these parameters are not universal for RDF. They may depend on the IMBH mass relative to the local disk mass, the radial dependence of the local disk mass, and the eccentricities. The particular scaling with the IMBH and SMBH mass and the Keplerian orbital time is motivated by Eq.~\eqref{eq:RDF_scaling}. Based on Eq.~\eqref{eq:inc}, the rate of reorientation for RDF increases exponentially as
\begin{equation}
\left.\frac{\rm{d}\iota}{\rm{d}t}\right|_{\rm RDF} =  [\lambda(\iota_0 - \iota_{\min}) - |\iota'_{\max}|] e^{\lambda t}\,,
\end{equation}
which may be expressed in terms of the instantaneous inclination of the IMBH in the form of Eq.~\eqref{eq:RDF_scaling} as
\begin{align}\label{eq:RDF-fit0}
\left.\frac{\rm{d}\iota}{\rm{d}t}\right|_{\rm RDF} &=  \lambda(\iota - \iota_{\min}) - |\iota'_{\max}|\nonumber\\
&= \left(
    \frac{8.2\iota - 3.28\iota_0 +1}{e^{0.636\,\iota_0}} -3.2\right) \frac{m_{\rm IMBH}}{m_{\rm SMBH}} t_{\rm orb}^{-1} \,.
\end{align}

This model may be interpreted physically as follows. Initially, the alignment rate is very slow, as the stellar distribution and potential are nearly axisymmetric for which $L_z$ is conserved. This is analogous to the lack of dynamical friction in a homogeneous medium before any particle made a close encounter and before a density wake develops. The gravitational influence of the IMBH warps the disk, which in turn torques the IMBH's orbit to align toward the disk. The speed of reorientation grows exponentially in the simulations, which is typical for other types of instabilities in galactic dynamics \citep[see, e.g.][]{Sellwood2013,Sellwood_Gerhard2020}. The instability here is driven by the IMBH, as the stellar disk was started from a stable configuration before deploying the IMBH. For larger $\iota_0$ the disk has more time to respond and becomes more warped by the time the IMBH aligns with the disk. The rate of alignment is largest when the orbit first reaches the edge of the warped disk; here, $\iota=\iota_{\min}$ and the first term vanishes in Eq.~\eqref{eq:RDF-fit0}. Since this configuration is roughly independent of the initial $\iota_0$, in terms of the inclination between the IMBH and the local disk, the final rate of alignment is roughly universal for fixed $(m_{\rm d,loc},m_{\rm IMBH},t_{\rm orb},\gamma)$. However, the bottleneck for alignment is in the initial phase, which is highly sensitive to $\iota_0$.

Up to this point, we have discussed the relative angle between the IMBH and the disk $\iota=\cos^{-1}(\hat{\bm{L}}_{\rm IMBH}\cdot \hat{\bm{L}}_{\rm disk})$. Let us now determine the rate of reorientation of the IMBH only, i.e., relative to an inertial frame. Since $\bm{L}_{\rm tot}=\bm{L}_{\rm IMBH}+\bm{L}_{\rm disk}$ is conserved, we expect that $(\rm{d} \iota_{\rm IMBH}/\rm{d}t)/(\rm{d} \iota/\rm{d}t) = [L_{\rm disk}/(L_{\rm disk}+L_{\rm IMBH})] = m_{\rm d,char} / (m_{\rm d,char} + m_{\rm IMBH}) $, where $m_{\rm d,char}=m_{\rm IMBH}L_{\rm disk}/L_{\rm IMBH}$ is the ``characteristic mass of the disk'' at the IMBH. Substituting in Eq.~\eqref{eq:RDF-fit0} gives
\begin{align}\label{eq:RDF-fit}
\left.\frac{\rm{d}\iota_{\rm IMBH}}{\rm{d}t}\right|_{\rm RDF} &=  \left(
    \frac{8.2\iota - 3.28\iota_0 +1}{e^{0.636\,\iota_0}} -3.2\right) \frac{\mu_{\rm IMBH}}{m_{\rm SMBH}} t_{\rm orb}^{-1} \,,
\end{align}
where
\begin{align}
    \mu_{\rm IMBH} &= \frac{m_{\rm IMBH}m_{\rm d,char}}{m_{\rm IMBH}+m_{\rm d,char}}\,,\\\label{eq:mdchar}
    m_{\rm d, char} &=\frac{L_{\rm disk}m_{\rm IMBH}}{L_{\rm IMBH}}
    =
    \frac{\int \rho r^{1/2} dV}{r^{1/2}} = 
    \frac{5 - 2\gamma}{4-2\gamma}\sqrt{\frac{r_d}{r}}m_{\rm d}\,.
\end{align}
Here, we substituted Eq.~\eqref{eq:rho} for $\rho$ with $\gamma<2$ in the limit of a thin flat Keplerian disk and an IMBH on a circular orbit at $r$. Note that $m_{\rm d, char} > m_{\rm d}$ as long as $r \leq r_{\rm d}$. Thus, if the IMBH is much less massive than the disk, $m_{\rm IMBH}\ll m_{\rm d}$, then $\mu_{\rm IMBH}\approx m_{\rm IMBH}$, as expected. 

Note that RDF is quite different from Chandrasekhar dynamical friction which is described by Eq.~\eqref{eq:talign} as
\begin{equation}\label{eq:CDF}
    \left.\frac{\rm{d}\iota_{\rm IMBH}}{\rm{d} t}\right|_{\rm CDF} = \frac{\ln \Lambda}{2\sin \iota \sin^3 (\iota/2)} \frac{m_{\rm IMBH}m_{\rm d,loc}}{m_{\rm SMBH}^2} t_{\rm orb}^{-1}\,.
\end{equation}

Comparing Eqs.~\eqref{eq:RDF-fit} and \eqref{eq:CDF}, it is clear that RDF is more efficient than CDF by a factor of $(m_{\rm SMBH}/m_{\rm d,loc})/[1+(m_{\rm IMBH}/m_{\rm d,char})]$, i.e. approximately $m_{\rm SMBH}/m_{\rm d,loc}$ for $m_{\rm IMBH}\ll m_{\rm d}< m_{\rm d,char}$, and a factor that depends on inclination. For our standard disk model, this mass factor is $440$. Thus, in comparison to CDF, the rate of alignment due to RDF is boosted most significantly for low-mass IMBHs (including heavy stellar-mass black holes), low-mass disks, and high-mass SMBHs. Note that for $(m_{\rm IMBH}, m_{\rm SMBH})=(10^3,10^{9})\,\Msun$ and with all other parameters unchanged, the rate of reorientation due to Chandrasekhar's dynamical friction and resonant dynamical friction are respective factors of $10^6$ and $10^3$ slower. Thus, in this case $t_{\rm CDF}/t_{\rm RDF} > 10^4$ for $\iota_0\sim 45^{\circ}$.
\begin{figure}
\includegraphics[width=1\columnwidth]{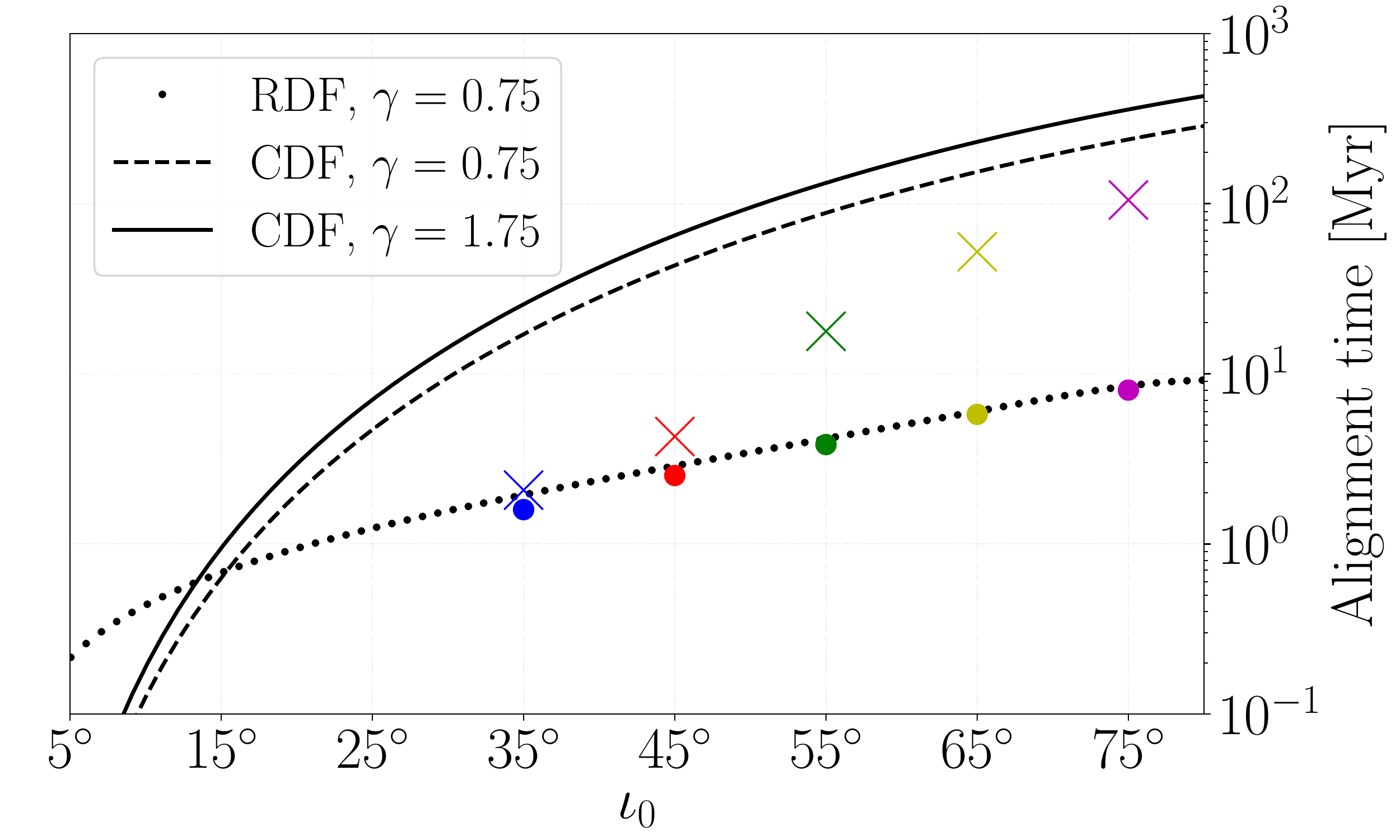}
\caption{\label{fig:empirical-fit} The alignment time of an IMBH via resonant dynamical friction measured in \textsc{N-ring} (symbols) versus the ordinary Chandrasekhar dynamical friction as a function of initial inclination for surface density exponent $\gamma=0.75$ (solid line) and $1.75$ (dashed line) (Eq.~\eqref{eq:talign}) for $\gamma=0.75$. Colored circles and $\times$ symbols show the results of the \textsc{N-RING} simulations for $\gamma=0.75$ and $1.75$, respectively. The dotted line shows the analytic fit to the simulations with $\gamma=0.75$ Eq.~\eqref{eq:RDF-fit0}. For $\gamma=1.75$, while the initial rate of alignment (not shown) is close to the dotted line, the  total alignment time is much longer for high $\iota_0$ because the relative inclination settles to an intermediate equilibrium for an extended period; see Figure~\ref{fig:inclination-gamma1.75}.}
\end{figure}

Figure \ref{fig:empirical-fit} compares the alignment time of an IMBH in \textsc{N-ring} for a disk of mass $m_{\rm d} = 8,191\Msun$ and that due to Chandrasekhar's dynamical friction (Eq.~\ref{eq:talign}) as a function of $\iota_0$. Dashed and solid black curves show the CDF (Eq.~\ref{eq:talign}) for surface density profile exponent $\gamma = 0.75$ and $1.75$, respectively, while the dotted black curve represents the empirical fit to RDF for $\gamma = 0.75$ (Eq.~\ref{eq:RDF-fit0}). Colored circles and crosses show the measured alignment times (up to the point when the relative inclination of the IMBH and the disk reaches $\iota = 15^\circ$) in the simulations with $\gamma = 0.75$ and $1.75$, respectively. These correspond to values where the IMBH crosses $15^\circ $ in Figure \ref{fig:inclination-t-long} and \ref{fig:inclination-gamma1.75} (top panel).
The figure shows that RDF becomes dominant over CDF in aligning the IMBH beyond a critical inclination
of $\iota_0 \gtrsim 15^\circ$. 

The analytic model (Eq.~\ref{eq:RDF-fit}) is successful at describing the \textit{initial} rate of reorientation for both models with $\gamma=0.75$ and $1.75$, respectively. However, it fails to describe the total alignment time for systems that exhibit the inclination hang-up phenomenon (e.g. $\iota_0\geq 55^{\circ}$ for $(N,m_{\rm IMBH},\gamma)=(8191,10^{3}\Msun,1.75)$). By comparing the disk structure for the different simulations shown in Figure~\ref{fig:diff-disk-particles}, we find that systems that exhibit the orbital hang-up phase form a discontinuity in the inclination distribution in a narrow range of radii in the outer disk; see upper right panel of Figure~\ref{fig:snapshots_Nring_175_i75} showing at 21 Myr in the region near $a=0.38\,\rm pc$. The analytic model presented here describes the initial rate of reorientation before discontinuities may form.

\section{Discussion} \label{sec:conclusions}
We have examined the dynamical evolution of a nuclear star cluster around an SMBH with both a spherical and a stellar disk component, as well as an IMBH on an initially inclined orbit with respect to the disk. We ran numerical simulations with two codes, $\varphi$\textsc{GPU} and \textsc{N-ring} with the same initial conditions. 

For an initial inclination of $\iota_0=45^{\circ}$ and for $(m_{\rm IMBH},m_{\rm SMBH} = (10^3\Msun, 10^6\Msun)$, we found the eccentricity of the IMBH decreased from the initial value of 0.33 to 0.02 in 4.5 Myr in $\varphi$\textsc{GPU}, while the eccentricity is fixed by construction in \textsc{N-ring}. This timescale for the IMBH's eccentricity decrease is much shorter than Chandrasekhar's dynamical friction timescale (Section~\ref{sec:nonresonant}), supporting previous findings by \citet{Madigan_Levin2012} that resonant dynamical friction, which arises due to orbit-averaged torques, dominates over ordinary dynamical friction driven by hyperbolic encounters and decreases the eccentricity for a disk corotating with the IMBH. This is related to the process of scalar resonant relaxation.

To investigate the contribution of VRR to how the orbital plane of the IMBH aligns with the stellar disk, we compared  $\varphi$\textsc{GPU} direct N-body simulations with the orbit-averaged \textsc{N-ring} simulations that follow VRR and neglect two-body relaxation and scalar resonant relaxation. The results of the two types of methods were in approximate agreement with respect to the evolution of angular momentum vector directions (Figure~\ref{fig:inclination-dependence}). Since VRR is weakly sensitive to the orbital eccentricity for $e<0.7$ if the spherical component of the cluster's potential drives rapid in-plane apsidal precession \citep{Kocsis2015}, the change of eccentricity does not influence the VRR evolution strongly. We have confirmed with $\varphi$\textsc{GPU} that the semimajor axis changes much more slowly during the alignment. The agreement between $\varphi$\textsc{GPU} and \textsc{N-ring} for the evolution of the orbital inclination of the IMBH and the disk suggests that the alignment is mainly driven by resonant gravitational torques between the stellar disk and the IMBH.

We investigated the response of the stellar disk due to the interaction with an IMBH. In the case where disk surface density follows $r^{-0.75}$, we found that the inner region of the disk exhibits a strong warp, while the outer disk is initially not affected (Figure~\ref{fig:snapshots}). Then, as the IMBH aligns with the midplane of the disk, the thickness of the disk increases in the radially overlapping regions and in the inner region. However, if the disk surface density is more cuspy and follows $r^{-1.75}$ consistent with the clockwise disk of massive stars in the Galactic Center \citep{Lu2009,Bartko2009,Bartko+2010,Yelda2014}, the evolution is qualitatively different, since in this case the mass is nearly uniformly distributed on a logarithmic scale, and there is much more mass in the inner disk and less mass in the outer disk in comparison to the shallower surface density profile. The disk is then stiffer in the inner region and more prone to deformations in the outer regions (Figure~\ref{fig:snapshots_Nring_175_i45} and \ref{fig:snapshots_Nring_175_i75}). In case of high initial IMBH inclination ($\iota \geq 45^\circ$) and if the fraction of IMBH mass to local disk mass is smaller than 0.7, then the IMBH alignment process is delayed and it takes place in three steps (Figure~\ref{fig:inclination-gamma1.75}): 
\begin{enumerate}[(i)]
    \item \label{i} The inclination decreases to an intermediate value of around  $\iota \sim 45^\circ$ on a timescale similar to the $\gamma=0.75$ case.
    \item \label{ii} The orbital inclination settles and oscillates with a decreasing amplitude for an extended period of time In this phase, while the inner disk perturbations are small, the outer disk ultimately responds in an irreversible way by developing a strong discontinuous warp starting near the outer edge of the disk, which propagates inward.
    \item \label{iii} Once the warp approaches the apoapsis of the IMBH orbit, the IMBH plunges rapidly into the disk. Ultimately, the inner disk is flat and aligned with the IMBH, and it becomes more warped and thicker in the outer regions. 
\end{enumerate}
We determined the alignment time of the IMBH with $\varphi$\textsc{GPU} and \textsc{N-ring} for 6 different initial inclinations ($\iota_0$) and found an analytic fit to the results. The evolution of the disk and the IMBH are in approximate agreement for $\varphi$\textsc{GPU} and \textsc{N-ring} if $\iota_0>20^{\circ}$.
The alignment time is an exponential function of $\iota_0$ if $\iota_0 < 90^\circ$ (Eq.~\ref{eq:RDF-fit} and Figure~\ref{fig:empirical-fit}) and there is no alignment in the counter-rotating case, i.e., if $\iota_0 > 90^\circ$. For the surface density profile of $r^{-0.75}$, the alignment of the IMBH with the stellar disk is between $9$--$28\times$ faster for $\iota_0 = 35^\circ$--$75^\circ$ than the estimate for the Chandrasekhar type dynamical friction (Eq.~\ref{eq:talign}). The orbital inclination alignment for $\iota>20^{\rm \circ}$  for component masses of $(m_{\rm SMBH},m_{\rm disk},m_{\rm IMBH})=(10^6\Msun,8\times 10^3\Msun,10^3\Msun)$ increases at an exponentially accelerating rate (Eq.~\ref{eq:inc}). These findings are valid for the total alignment time for surface density $r^{-0.75}$ for a wide range of masses and inclinations, and also for $r^{-1.75}$ surface density as long as the initial inclination satisfies $\iota_0\lesssim 45^{\circ}$ and the IMBH mass is not much smaller than the local disk mass. However, for surface density $r^{-1.75}$ with large $\iota_0$ and/or small IMBH mass, the exponential fit to the realignment (Eq.~\ref{eq:inc}) is valid approximately only for phase (\ref{i}), but the total alignment is delayed due to  the orbital inclination hang-up phase (\ref{ii}). Nevertheless, this is typically still much faster than the Chandrasekhar's dynamical friction (Figure~\ref{fig:empirical-fit}). The time duration of the inclination hang-up phase increases with the smoothness of the disk by increasing the number of disk particles for fixed total disk mass (Figure \ref{fig:diff-disk-particles}).

We also examined the dependence of the alignment rate on the mass of the IMBH and found that it is initially linearly proportional to mass, similarly to Chandrasekhar's dynamical friction, if the IMBH mass is much larger than the mass of disk particles (Figure~\ref{fig:logiperlogm}). At later times, as the disk gets significantly perturbed, the scaling with $m_{\rm IMBH}$ can be either stronger or weaker, depending on the surface density exponent $\gamma$ and $m_{\rm IMBH}$. Furthermore, we found that the resonant dynamical friction timescale is approximately proportional to $m_{\rm SMBH}/m_{\rm IMBH}$ unlike Chandrasekhar's dynamical friction, which scales with $m_{\rm SMBH}^2/(m_{\rm d}m_{\rm IMBH})$. This implies that, while RDF is faster than CDF by a factor of 10--40 for our fiducial model with $(m_{\rm IMBH},m_{\rm SMBH})=(10^3\Msun,10^6\Msun)$, RDF is faster than CDF by a factor of more than $\sim 10^4$ for $(m_{\rm IMBH},m_{\rm SMBH})=(10^3\Msun,10^9\Msun)$.

These findings may provide an explanation for the dynamical origin of the warped structures of the stellar disks in the Galactic Center \citep{Bartko2009}. Our results indicate that an inclined massive perturber, like an IMBH, can strongly warp an initially flat stellar distribution, due to coherent gravitational torques. Depending on the mass of the IMBH and the surface density profile of the disk, it may cause a strongly perturbed inner or outer disk and a significantly thickened overlapping disk on a $1-100$ Myr timescale after its formation or its arrival to the Galactic Center, see e.g., Figure \ref{fig:snapshots}. 
\begin{figure*}[t!]
\includegraphics[width=1\textwidth]{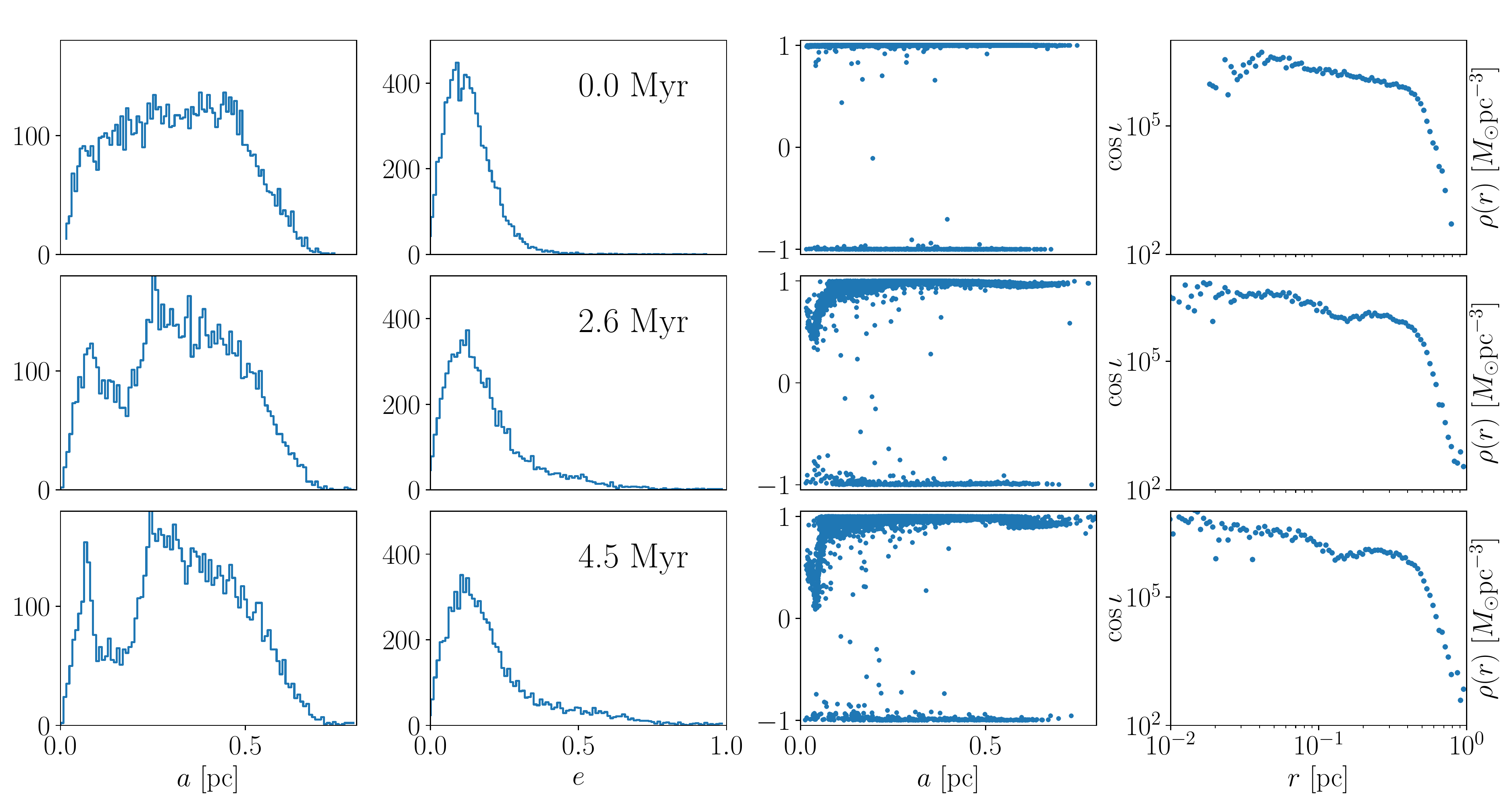}
\caption{\label{fig:obital-elements} Orbital parameter distributions of stars in the disk: the number density of semimajor axes and eccentricities, the distribution of inclinations as a function semimajor axis, and the surface density of the disk as a function of radius, respectively, from left to the right. First row shows the initial conditions at $0$ Myr of the $\varphi$\textsc{GPU} simulation, second row shows the parameters at $2.6$ Myr, and the third row shows parameters at $4.5$ Myr.}
\end{figure*}

Resonant dynamical friction may have applications beyond the relaxation of IMBHs examined in this paper. It may affect all objects in stellar clusters much more massive than the individual constituents of the disk, if present, including massive stars, stellar mass black holes (BHs), or the center of mass of massive binaries. Furthermore, it is also expected to operate in any type of disk with a high number of particles, including active galactic nucleus (AGN) accretion disks. Previously, it has been argued that stars and BHs crossing the disk on low-inclination orbits get captured by Chandrasekhar dynamical friction into the disk \citep{Bartos2017,Panamarev2018,Tagawa2020}. An interesting implication is that, if BHs settle into the disk, they interact dynamically and form BH-BH binaries efficiently, and frequent dynamical interactions and gas effects drive the BHs to merger, producing gravitational waves (GWs)  detectable by LIGO, VIRGO, and KAGRA \citep{McKernan+2014,McKernan+2018,Bartos2017,Leigh2018,Yang+2019,Tagawa2020,Tagawa2021,Samsing2020}. Mergers are also facilitated by Lidov-Kozai oscillations in anistropic systems \citep{Heisler_Tremaine1986,Petrovich_Antonini2017,Hamilton_Rafikov2019}. The results in this paper show that resonant dynamical friction may accelerate the capture of objects in the accretion disks by a factor proportional to the SMBH mass over the local disk mass for large orbital inclinations. Pressure and viscosity in a gaseous disk do not inhibit the orbit-averaged torque from the IMBH, which leads to realignment and the warping of the disk \citep{Bregman_Alexander2012}. Thus, RDF may efficiently catalyze the alignment of the orbital planes of BHs even in low-luminosity AGN or Seyfert galaxies with relatively small disk masses, which may not be possible for Chandrasekhar dynamical friction. In fact, this mechanism extends the scope of the ``AGN merger channel'' for GW source populations even beyond low-luminosity AGN and Seyfert galaxies, as it may organize BHs into disks also in nonactive galaxies with nuclear stellar disks. 

Even more generally, resonant dynamical friction may be expected to operate in all systems whose mean-field potential admits action-angle variables where one of the precession frequencies is zero, e.g. for approximately spherical systems. RDF may accelerate the relaxation of massive subsystems including dark matter substructures in the spherical halo into galactic disks, which could then catalyze the formation of galactic spiral arms \citep{DOnghia+2013}, similarly to how an IMBH may excite spiral waves in a stellar disk in the Galactic center \citep{Perets+2018}. It may also play a role in the formation and evolution of anisotropic groups of galaxies, including the disk of Milky Way satellites \citep{Kroupa+2005} or anisotropic galaxy clusters \citep{Binney1977}.
\begin{figure*}[t!]
\includegraphics[width=1\textwidth]{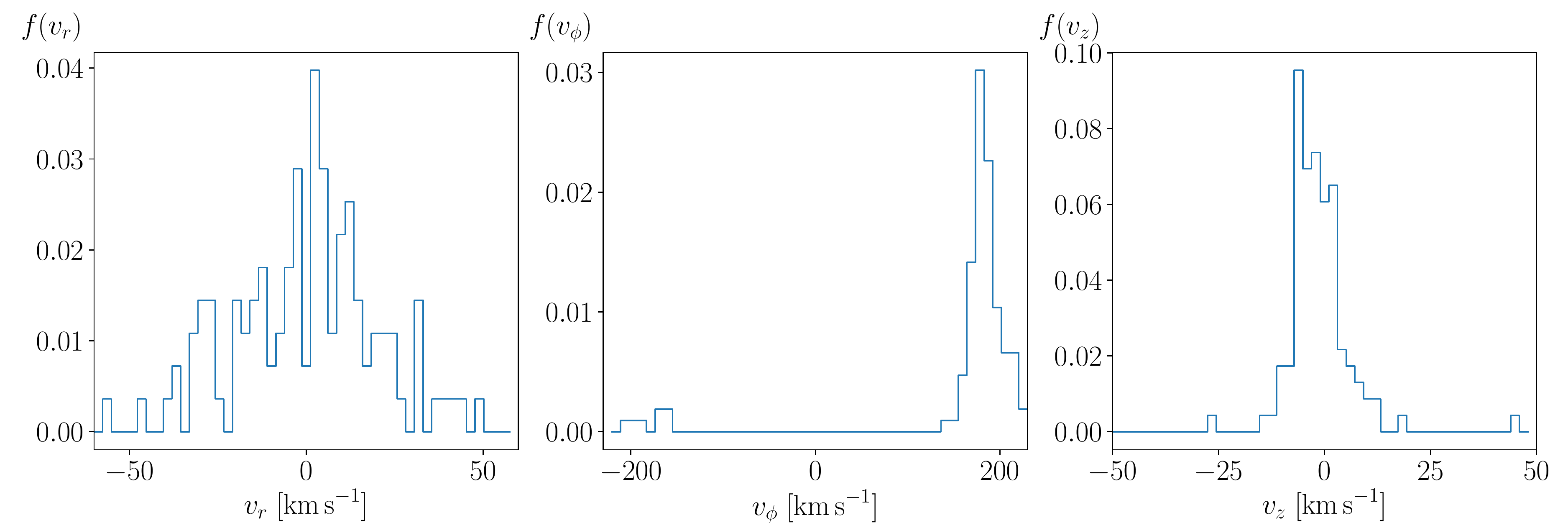}
\caption{\label{fig:velocity-distribution}Velocity density histograms of stars in the vicinity of the IMBH's crossing point of the stellar disk during the first crossing.}
\end{figure*}

\acknowledgments{
We gratefully thank Peter Berczik for providing us with the $\varphi$\textsc{GPU} code and for his help in its use. We thank Scott Tremaine, Yohai Meiron, Hiromichi Tagawa, Nathan Leigh, Chris Hamilton, and John Magorrian for useful discussions. We are grateful to M\'aria Kolozsv\'ari for help with logistics and administration related to the research. This work received funding from the European Research Council (ERC) under the European Union's Horizon 2020 research and innovation program under grant agreement No 638435 (GalNUC) and was supported by the Hungarian National Research, Development, and Innovation Office grant NKFIH KH-125675. \'Akos Sz\"olgy\'en was supported by the \'UNKP-20-3 New National Excellence Program of the Ministry for Innovation and Technology from the source of the National Research, Development and Innovation Fund.}

%\appendix
\section*{Appendix}

\begin{center}
    \textit{Distribution of orbital parameters}
\end{center}

\label{sec:appendix}
Here, we give the semimajor-axis, eccentricity, inclination, and mass distributions of this paper's fiducial relaxed stellar disk model that we introduced in Section~\ref{sec:initial}. Figure \ref{fig:obital-elements} shows the number density of semimajor axes and eccentricities, the distribution of the cosine of the inclinations as a function semimajor axes, and the surface density as a function of radius ($\propto r^{-0.75}$), respectively, in panels from the left to the right. The first row of panels represents the initial, relaxed disk model at $0$ Myr when the IMBH is added to the simulation. The second and the third row show snapshots of $\varphi$\textsc{GPU} simulation at $2.6$ Myr (when the IMBH reaches the disk), and $4.5$ Myr (end of the simulation). Here, the initial IMBH parameters are $(m_\mathrm{IMBH},$ $a_\mathrm{IMBH},e_\mathrm{IMBH},\iota_0) = (10^3\Msun,0.15\mathrm{pc},0.33,45^\circ)$.

The first column of panels show the evolution of disk stars' semimajor axes. As the IMBH's orbital inclination enters the disk at $2.6$ Myr, the number density of stars around the IMBH's semimajor axis ($0.15\, \rm pc$) is strongly depleted. By $4.5$ Myr, the number density decreases by a factor of $\sim50\%$ which suggests that the IMBH strongly scatters the disk stars out of their orbit, i.e., nonresonant dynamical friction becomes dominant over VRR.

The eccentricity density of stars does not vary much during the evolution, as shown by the second column of panels. However, the eccentricity distribution flattens as the peak decreases by $22\%$ and the standard deviation increases by $50\%$. Note that, unlike \citet{Gualandris+2012}, we do not find the eccentric disk instability that forms a bimodal eccentricity distribution, but this may be due to the differences in our assumptions. The disk is initially axisymmetric in our simulations, and we account for the spherical potential of the stellar cusp, which drives differential apsidal precession and quenches scalar resonant relaxation.

The inclination of stars as a function of the semimajor axis shows a warp and an increase of the thickness of the disk in the third column of panels, which is also presented in Figure \ref{fig:snapshots} and \ref{fig:inclination-t} in detail.

Figure \ref{fig:velocity-distribution} shows the local 3D velocity density of stars in cylindrical coordinates at $0$ Myr. The velocity distribution $f(v_r,v_\phi,v_z)$ is restricted to the stars in the neighborhood of the crossing point of the IMBH in the disk in a cylindrical box with radius $r_\mathrm{cyl} = 0.05$ pc and height $h_\mathrm{d} = 0.01$ pc.

\bibliography{szolgyen_et_al}

\end{document}